\documentclass[prb,showpacs,twocolumn,groupedaddress,superscriptaddress]{revtex4-2}

\usepackage{graphicx,graphics,color,epsfig}
\usepackage{amsmath}
\usepackage{amssymb}
\usepackage{amsfonts}
\usepackage{amsfonts}
\usepackage{epstopdf}
\usepackage{bm}
\usepackage{color}
\usepackage{caption}
\usepackage{subcaption}

\begin{document}

\preprint{APS/123-QED}

\title{Internal consistency of multi-tier $GW$+EDMFT}

\author{Ruslan Mushkaev}
\email{ruslan.mushkaev@unifr.ch}
\affiliation{Department of Physics, University of Fribourg, 1700 Fribourg, Switzerland}
\author{Francesco Petocchi}%
\affiliation{Department of Quantum Matter Physics, University of Geneva, 1211 Geneva 4, Switzerland}
\author{Viktor Christiansson}%
\affiliation{Department of Physics, University of Fribourg, 1700 Fribourg, Switzerland}
\author{Philipp Werner}%
\affiliation{Department of Physics, University of Fribourg, 1700 Fribourg, Switzerland}

\date{\today}

\begin{abstract}
The multi-tier $GW$+EDMFT scheme is an {\it ab-initio} method for calculating the electronic structure of correlated materials. While the approach is free from {\it ad-hoc} parameters, it requires a selection of appropriate energy windows for describing low-energy and strongly correlated physics. In this study, we test the consistency of the multi-tier description by considering different low-energy windows for a series of cubic SrXO$_3$ (X=V,Cr,Mn) perovskites. Specifically, we compare the 3-orbital $t_{2g}$ model, the 5-orbital $t_{2g}+e_g$ model, the 12-orbital $t_{2g}+O_p$ model, and (in the case of SrVO$_3$) the 14-orbital $t_{2g}+e_g+O_p$ model
and compare the results to available photoemission and X-ray absorption measurements. 
The multi-tier method yields consistent results for the $t_{2g}$ and $t_{2g}+e_g$ low-energy windows, while the models with $O_p$ states produce stronger correlation effects and mostly agree well with experiment, especially in the unoccupied part of the spectrum. 

We also discuss the consistency between the fermionic and bosonic spectral functions and the physical origin of satellite features, and present momentum-resolved charge susceptibilities.
\end{abstract}

\maketitle

\section{\label{intro}Introduction}

Materials with strong electron-electron interactions exhibit diverse and remarkable properties such as high-$T_c$ superconductivity, colossal magnetoresitance, and interaction driven metal-insulator transitions, and are thus of central interest in condensed matter physics \cite{Dagotto2005}. A widely studied subclass of these materials are the transition metal oxides, which contain partially filled narrow $d$-electron bands. These give rise to strong correlation phenomena, which cannot be captured at the level of effective single-particle descriptions. Theoretical studies of transition metal oxides often involve a combination of Density Functional Theory (DFT) \cite{Hohenberg1964,Kohn1965} in the Local Density Approximation (LDA) \cite{Kohn1965} and Dynamical Mean Field Theory (DMFT) \cite{Georges1996}. The combined DFT+DMFT scheme has been successful in reproducing a wide range of correlated electron phenomena, including mass enhancements \cite{HauleDFT_DMFT}, and localized to itinerant electronic transitions \cite{haule_science}. This approach however suffers from two main drawbacks: the reliance on {\it ad-hoc} parameters such as Hubbard $U$ and Hund's $J$ interactions, and the double counting of interaction energies \cite{KarolakDC_DFT}. DFT+DMFT hence cannot be considered a truly {\it ab-initio} method, and more sophisticated approaches need to be developed for reliable first-principles predictions of the properties of correlated materials. 

\begin{figure*}[ht!]
    \includegraphics[width=0.9\textwidth]{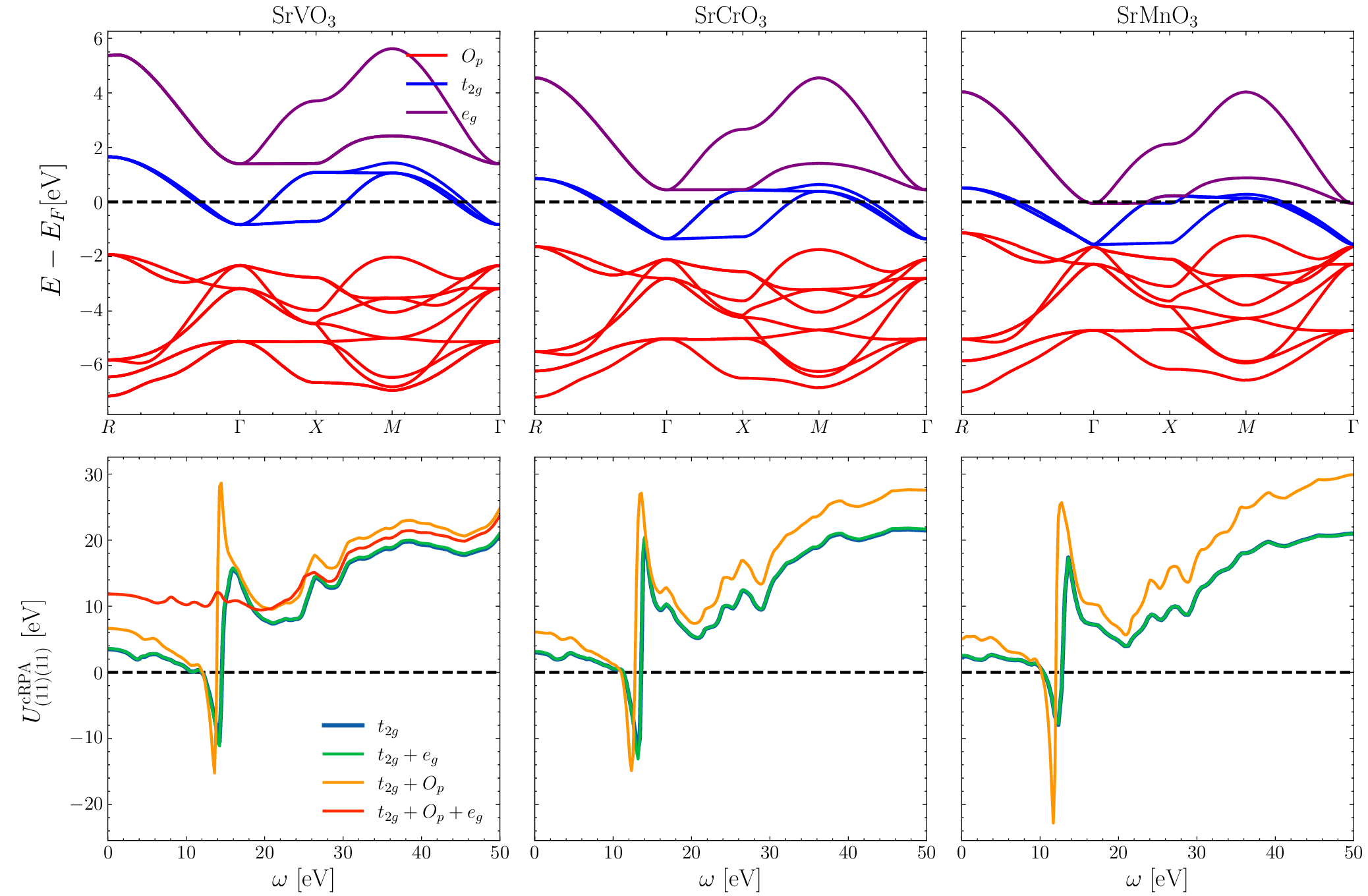}
    \caption{Top row: DFT-LDA bandstructure of the cubic perovskite SrXO$_3$ family. The three $t_{2g}$ and two $e_g$ bands originate from the X cation's $d$-orbitals in the cubic environment. The 9 lower lying bands originate from Oxygen $p$-states. Bottom row: real part of the intra-orbital cRPA Coulomb interaction $U$ in the $t_{2g}$ manifold for the cubic perovskites. 
    The high-frequency values of these curves correspond to the bare Coulomb interactions. The pole feature at $\omega \sim 13 \ \mathrm{eV}$ in the 3-, 5- and 12-band models, which is absent in the 14-band model (see SrVO$_3$ results) can be attributed to $O_p$-$e_g$ screening processes.}
    \label{dft_bands_crpa}
\end{figure*}

A promising way forward is the $GW$ + Extended Dynamical Mean Field Theory (EDMFT) \cite{Biermann2003,Boehnke2016,NilssonGW_EDMFT,Choi2023} method which is based on the $GW$ formalism \cite{HedinGW} and supplements the nonlocal $GW$ self-energy and polarization with more accurate estimates of the local components calculated by EDMFT. Since both $GW$ and EDMFT are diagrammatic methods, a double counting of self-energy and polarization contributions can be easily avoided. A further advantage of the $GW$+EDMFT formalism is that it involves no {\it ad-hoc} interaction parameters, since these are computed self-consistently, making it a genuinely {\it ab-initio} method. 
In the multi-tier implementation of the framework \cite{Boehnke2016,NilssonGW_EDMFT}, different degrees of freedom are treated at an appropriately chosen level of approximation. The full band space (TIER III) is treated at the level of one-shot $GW$ ($G^0 W^0$), while the physically more relevant low-energy subspace around the Fermi level (TIER II) is solved with the fully self-consistent $GW$ + EDMFT scheme. In this self-consistency loop, due to the complexity of the many-body problem, only the local correlations of the most strongly correlated orbitals (TIER I) are treated with EDMFT.  

If the downfolding from the higher-energy to the lower-energy subspaces was implemented exactly,  and we could exactly solve the resulting many-body problem, then the low-energy correlated electronic structure would be independent of the choice of energy windows.

In reality, approximations are made, both in the downfolding \cite{Honerkamp2018} and in the solution of the low-energy effective theory (e.~g. density-density approximation in the impurity interaction), and it is thus important to understand how much the low-energy physics depends on the choice of the different tiers in the multi-tier approach. 
In this study, we systematically test the internal consistency of our $GW$+EDMFT scheme using the well characterized and studied family of correlated perovskite compounds SrXO$_3$ (X=V, Cr, Mn) as our test materials. Specifically, we will compare the 3-band ($t_{2g}$) description to the five-band ($t_{2g}+e_g$) modeling and to a 12-band treatment, which in addition to the $t_{2g}$ states also includes the Oxygen $p$ manifold. In the case of SrVO$_3$, we will also consider the 14-band $t_{2g}+O_p+e_g$ model.

This paper is structured as follows: Section \ref{results} presents the band structures and effective interactions for the tier II calculations, and the results including spectral functions and charge susceptibilities. Section \ref{disc} discusses the most important findings, while Section~\ref{method} contains a brief outline of the method. A more detailed description of the multi-tier approach and of the susceptibility calculations can be found in the Supplemental Material.

\section{\label{results} Results}

\subsection{\label{input} DFT bandstructure and model spaces}

The first step in generating the input for the $GW$+EDMFT self-consistency is to compute the DFT bandstructure of the ``full" space, which contains correlation effects in the form of the exchange correlation potential. This bandstructure allows us to identify a relevant low-energy sector around the Fermi energy to which we can downfold, for use in the subsequent $GW$+EDMFT self-consistent calculations.

\begin{figure}[ht!]
    \begin{centering}
\includegraphics[width=0.4\textwidth]{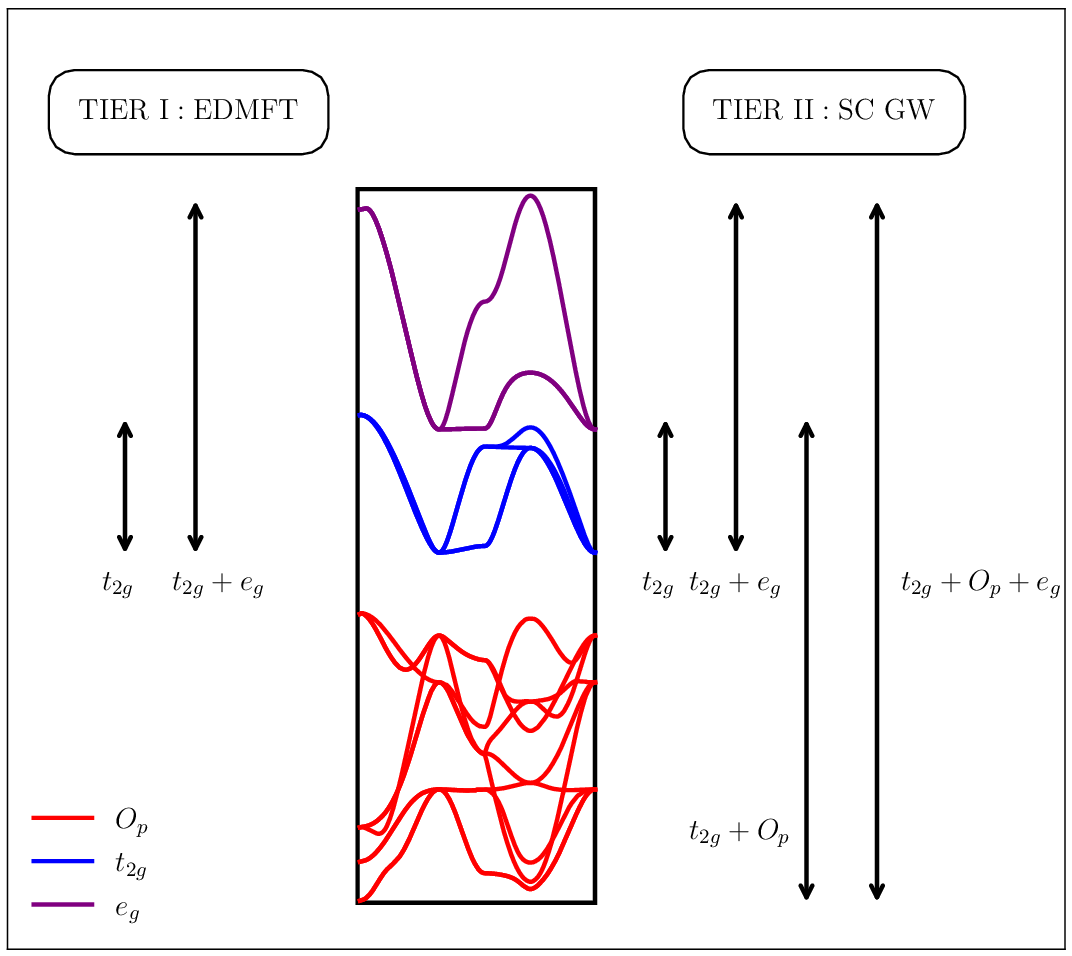}
\end{centering}
    \caption{Different model spaces used in the $GW$+EDMFT simulations of the cubic perovskites. The most correlated $t_{2g}$ and $t_{2g}+e_g$ local subspaces are treated with EDMFT, while their non-local parts, as well as the lower lying Oxygen $p$-states are treated by self-consistent $GW$.}
    \label{model_spaces}
\end{figure}

In the cubic perovskites SrXO$_3$ (where X=V, Cr, Mn), the X cation is enclosed in an Oxygen octahedron, which is in turn enclosed in a Strontium cube. The most relevant low-energy bands around the Fermi level have predominantly $t_{2g}$ character and originate from the X cation's $3d$ orbitals, see the blue bands in Fig. \ref{dft_bands_crpa}. The two higher lying bands are $e_g$-like and are crystal-field-split from the $t_{2g}$. The 9 lower lying bands originate mainly from the Oxygen $p$ states. At the DFT level, Vanadium, Chromium and Manganese host 1, 2 and 3 electrons in the $t_{2g}$ bands,  respectively. For our tests of the consistency of the method, we will consider the $t_{2g}$, $t_{2g}+e_g$ and $t_{2g}+O_p$ model spaces (as well as $t_{2g}+O_p+e_g$ for SrVO$_3$), referring to the orbitals contained in the self-consistent (TIER II) subspace, see Fig.~\ref{model_spaces}. For the TIER I subspace, which is treated with EDMFT, we choose the subspace of $d$ orbitals, where we expect the strongest correlation effects.

\subsection{SrVO$_3$}

SrVO$_3$ is a well-studied \cite{Morikawa1995,Sekiyama2004,Pavarini2004,srvo3_prl,Sakuma2013,TomczakSrvo3,BackesSrvo3,Boehnke2016,Nakamura2016,PetocchiScreening} metallic perovskite compound, with a near-perfect cubic structure. Experimentally, it is found to be a metal with moderately renormalized conduction bands. Near the Fermi level, photoemission spectroscopy measurements \cite{srvo3_prl} find a reduction of the band velocity by about a factor of two, compared to the LDA bandstructure. The photoemission spectra furthermore reveal an almost dispersionless feature around $-1.5$~eV, which has been interpreted as a lower Hubbard band \cite{Pavarini2004,Sekiyama2004}, or as a signal originating primarily from Oxygen vacancies \cite{BackesSrvo3}. Inverse photoemission \cite{TomczakSrvo3} and X-ray absorption spectra \cite{Inoue1994} reveal a broad peak centered at 2.5-3 eV in the unoccupied part of the spectrum, which could be an upper Hubbard band \cite{Pavarini2004,Sekiyama2004}, a plasmon sideband \cite{Boehnke2016,Nakamura2016,PetocchiScreening}, or a feature originating from $e_g$ states \cite{TomczakSrvo3}. 

\subsubsection{Filling, interactions, and mass enhancements}

At the LDA level, SrVO$_3$ contains one electron per unit cell in the $t_{2g}$ conduction bands (V 3$d_{xy,xz,yz}$ orbitals). Its $t_{2g}$ bands are isolated from the empty $e_g$ bands and the filled Oxygen $p$ bands (see Fig.~\ref{dft_bands_crpa}).

\begin{figure*}[ht!]
    \centering
    \includegraphics[width=0.9\textwidth]{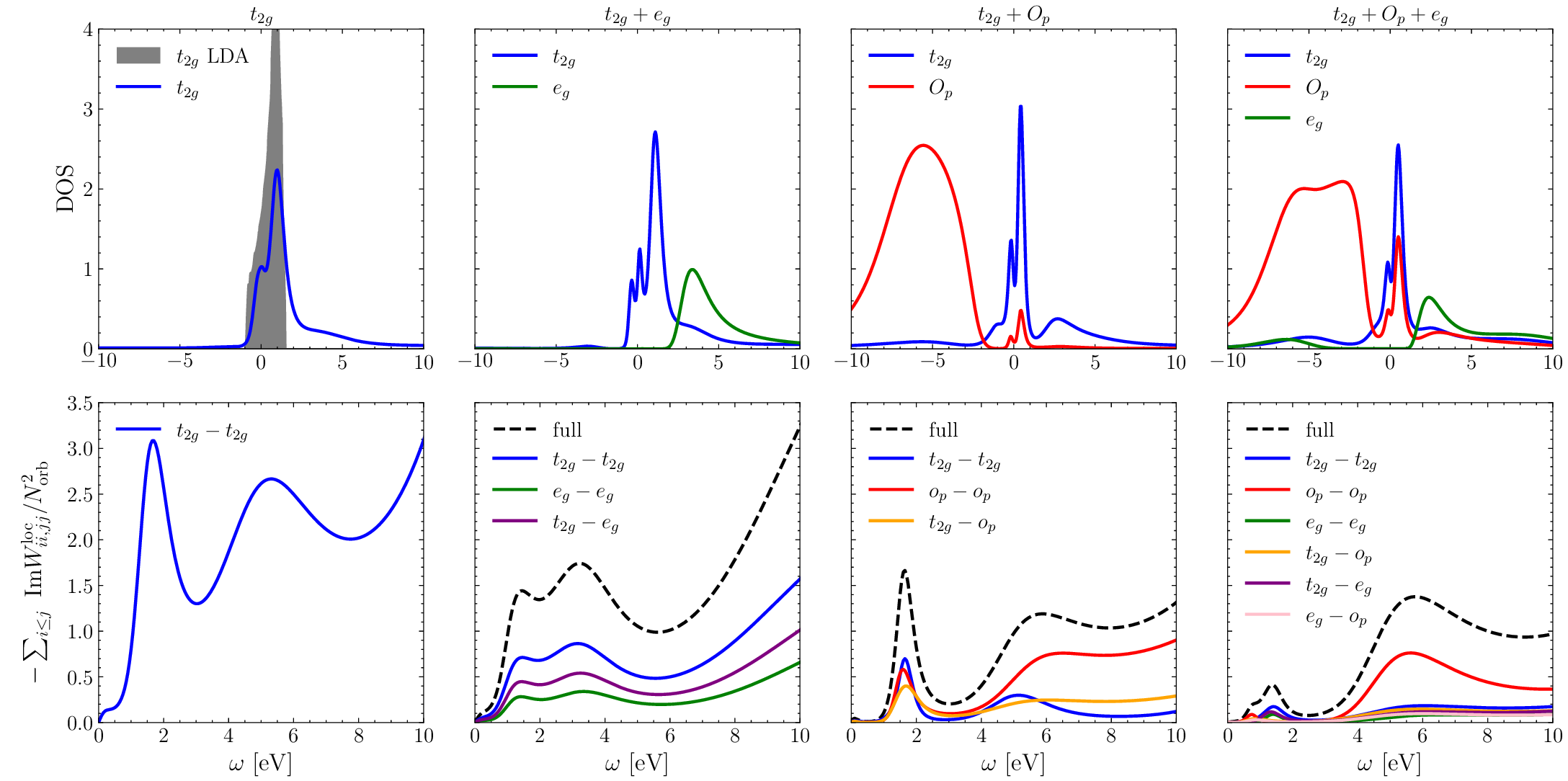}
    \caption{Top row: local spectral functions of SrVO$_3$ computed within different self-consistent model spaces. All spectra display a central quasiparticle peak along with satellite features in the $t_{2g}$ sector. Bottom row: full and orbital-resolved local screened interactions computed as a sum of a subset of density-density matrix elements.
    }
    \label{dosloc_wloc_srvo3}
\end{figure*}

In the $GW$+EDMFT results, shown in Fig.~\ref{dosloc_wloc_srvo3}, the local $t_{2g}$ DOS  displays a prominent quasiparticle peak, moderately renormalized with respect to the LDA DOS, and two side peaks which are not captured by DFT -- a weak satellite in the range from $-3$ to $-1$ eV and a more prominent broad feature around $2.5$-$3$ eV. To test the consistency of the $GW+$EMDFT results, we take a closer look at the evolution of the spectral features with increasing size of the self-consistent (SC = Tier II + Tier I) space, and check how they compare with experimental data.  

\begin{table}[t]
    \caption{$t_{2g}$ filling, quasiparticle weight, atomic gap, local static interactions (cRPA $U$ and self-consistently calculated $\mathcal{U}$ and $\mathcal{J}$), and mass enhancement ratio for SrVO$_3$ and the indicated model spaces.
    }
    \label{srvo3_table}
    \centering
    \begin{tabular}{ |p{1.8cm}|p{1.5cm}|p{1.5cm}|p{1.5cm}|p{1.66cm}| }
    \hline
      & $t_{2g}$ & $t_{2g}+e_g$ & $t_{2g}+O_p$ & $t_{2g}+O_p+e_g$  \\
     \hline
     \hline
    $N_{t_{2g}}$
    & 1.00 & 0.87 & 1.91 & 1.93\\
     \hline
     $Z_{t_{2g}}^\text{EDMFT}$ & 0.61   & 0.64 & 0.36 & 0.44 \rule[-1.5ex]{0pt}{4ex} \\
     \hline
     $\Delta_{\mathrm{at}}  \ [\mathrm{eV}] $ & 1.5   & 1.6 & 3.6 & 7.7 \\
     \hline
     $\mathcal{U}^{\mathrm{static}}_{t_{2g}}  \ [\mathrm{eV}] $ & 2.7   & 2.8 & 5.1 & 9.2\\
     \hline
     $\mathcal{J}^{\mathrm{static}}_{t_{2g}}  \ [\mathrm{eV}] $ & 0.4   & 0.4 & 0.5 & 0.5\\
     \hline
     $U^{\mathrm{cRPA}}_{t_{2g}}  \ [\mathrm{eV}] $ & 3.5   & 3.6 & 6.6 & 11.8 \\
     \hline
     $\frac{m_{\mathrm{GW+EDMFT}}}{m_{\mathrm{LDA}}}$ & 1.2   & 1.0 & 1.5 & 1.6 \rule[-1.5ex]{0pt}{4ex} \\
     \hline
    \end{tabular} 
\end{table}

\begin{figure*}[ht!]
    \centering
    \includegraphics[width=0.98\textwidth]{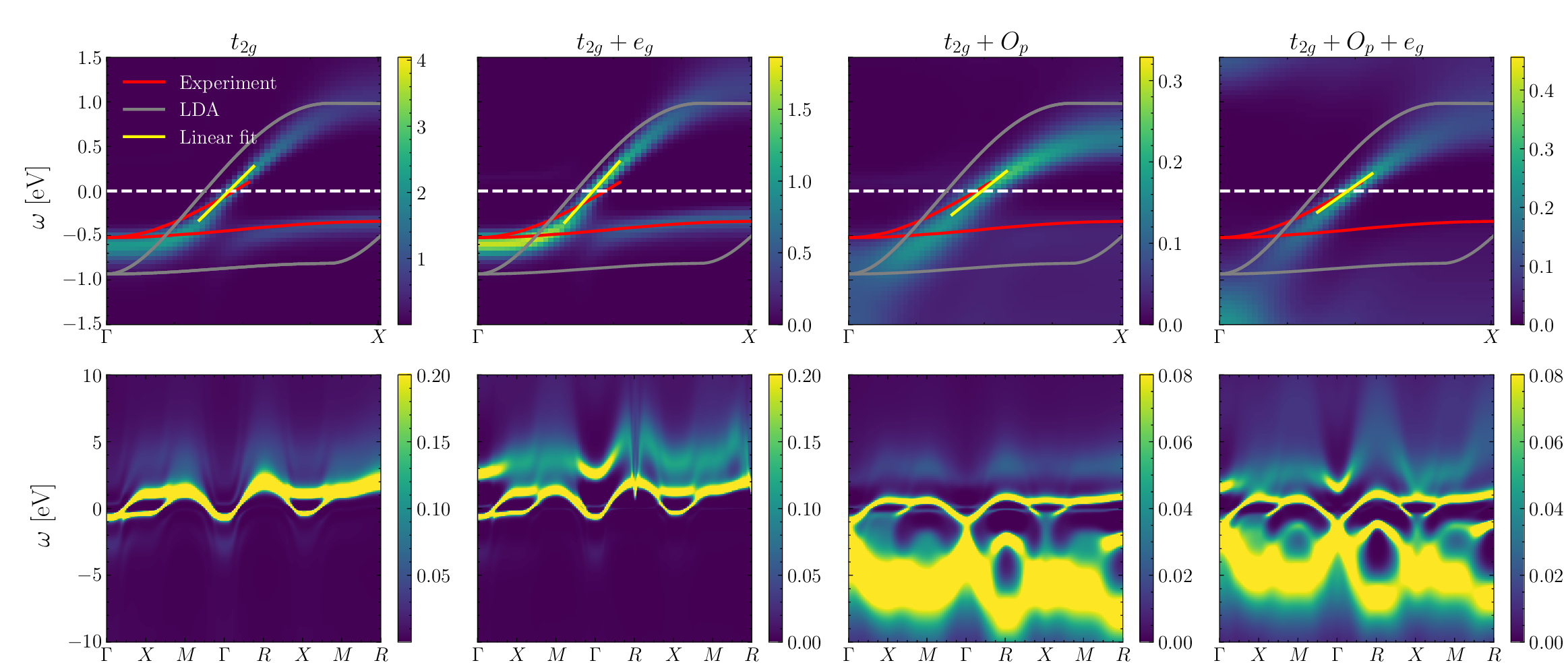}
    \caption{Top row: momentum resolved correlated bandstructure along the indicated path, and comparison to the peak intensity of the momentum resolved photoemission spectrum reported in Ref.~\onlinecite{BackesSrvo3} (red lines). The mass renormalizations (with respect to the LDA values) in Tab.~\ref{srvo3_table} are extracted from linear fits of the Fermi velocity (yellow lines). Bottom row: momentum resolved spectral function on the full path in the Brillouin zone.}
    \label{srvo3_akw_exp}
\end{figure*}

The last column of Tab.~\ref{srvo3_table} reports the $GW$+EDMFT mass enhancements with respect to the LDA bandstructure, computed as a ratio of Fermi velocities. The corresponding fits to the correlated bands are shown in Fig.~\ref{srvo3_akw_exp}. We note that while the mass enhancements of $1.0$-$1.2$ are comparable in the $t_{2g}$ and $t_{2g}+e_g$ models, up to the numerical uncertainty of the fitting procedure, the $t_{2g}+O_p$ and $t_{2g}+O_p+e_g$ models yield a stronger mass enhancement of $1.5-1.6$, which is closer to the experimental value of $m_{\mathrm{exp}}/m_{\mathrm{LDA}} \sim 2$. \cite{srvo3_prl}

A larger SC space implies an increased degree of localization of the Wannier functions spanning this space, and a larger number of screening channels which are excluded in the cRPA construction. As a result, both the bare and screened interactions are larger in the models including $O_p$, compared to the $t_{2g}$ and $t_{2g}+e_{g}$ models (see Fig.~\ref{dft_bands_crpa}). We list the static ($\omega=0$) values of the local cRPA and impurity interactions in Tab.~\ref{srvo3_table}. 
Without the oxygen bands in the model, adding the $e_g$ states to the $t_{2g}$ manifold has little effect on the effective $t_{2g}$ interaction, consistent with previous findings \cite{PetocchiScreening}. The impact of the $e_g$ states however increases drastically when we further include the Oxygen states, as evidenced by a significant enhancement of the impurity interaction in the $t_{2g}+O_p + e_g$ model when compared to the $t_{2g}+O_p$ one. This demonstrates the important role of the $e_g$-$O_p$ screening channels.

For the $t_{2g}$ model, our cRPA static interaction is close to the value of $U$ reported in Ref.~\onlinecite{Nakamura2016}, while the static effective impurity interaction $\mathcal{U}^\text{static}$ is almost identical to the difference $U-V$ (with $V$ the nearest-neighbor interaction), and hence to the naive estimate of the effective local interaction which takes nonlocal screening into account \cite{Schueler2013}. This confirms that our $GW$+EDMFT framework yields meaningful effective interactions.

A second effect, which influences the degree of correlations in our calculations, is the filling of the $t_{2g}$ orbitals (indicated by $N_{t_{2g}}$ in Tab.~\ref{srvo3_table}). Due to the hybridization with Oxygen $p$ states, $N_{t_{2g}}$ in the $t_{2g}+O_p$ and $t_{2g}+e_g+O_p$ models (henceforth referred to as ``$O_p$ models") is almost twice the filling of the three band counterpart.
The ``extra" electron in these models can be found by integrating the $t_{2g}$ local spectral function of Fig.~\ref{dosloc_wloc_srvo3} in an energy window covering the Oxygen states.

\subsubsection{Fermionic spectral functions}

\begin{figure}[t]
    \centering
    \includegraphics[width=0.45\textwidth]{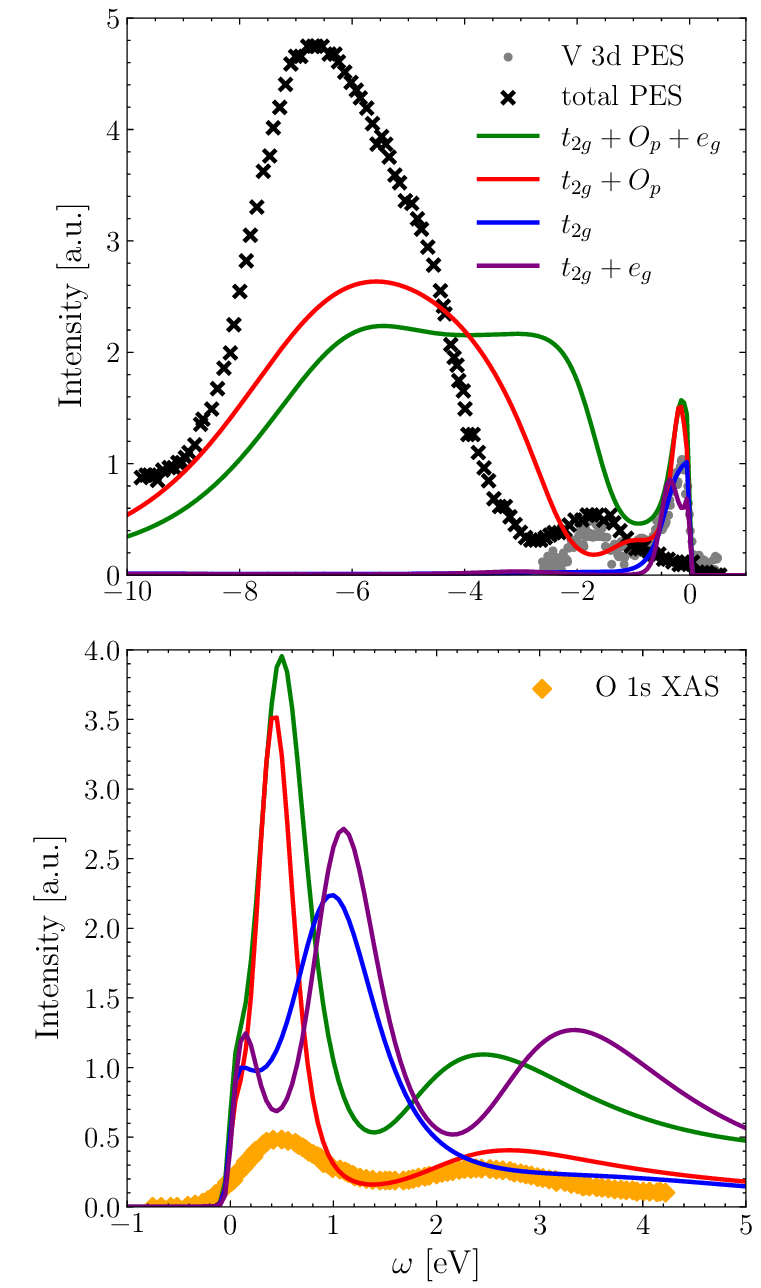}
    \caption{Comparison of local spectral functions to experimental PES/XAS spectra from Ref.~\onlinecite{Inoue1994}. The total and V 3d PES have been measured at $T=80$~K and $T=20$~K respectively. The calculated spectral functions have been multiplied by the Fermi-Dirac function $f(\omega)$ for $T=20$~K and by $1-f(\omega)$ for $T=298$~K in the occupied and unoccupied regions, respectively.
    }
    \label{srvo3_pes_xas_dos}
\end{figure}

A comparison between the occupied parts of the local spectral function and photoemission spectroscopy (PES) data from Ref.~\onlinecite{Inoue1994} is shown in the top panel of Fig.~\ref{srvo3_pes_xas_dos}. We note that the occupied part of the quasiparticle peak is well reproduced by all models, and that the $t_{2g}+O_p$ spectral function appears to best capture a portion of the incoherent spectral weight (the latter is very broad and weak at $\sim -3 $ eV in the $t_{2g}$ and $t_{2g}+e_g$ models).
There is a widespread view in the literature that the feature found in experiments at $- 1.5$ eV represents a lower Hubbard band \cite{Pavarini2004,Sekiyama2004}. Its relative amplitude with respect to the quasiparticle peak might however be severely overestimated by the presence of Oxygen vacancies generated by high UV doses \cite{BackesSrvo3}. The latter produce states at the same energy as the purported lower Hubbard band. 

When compared to experimental PES, the $O_p$ models underestimate the binding energy of the Oxygen states by roughly 1-2 eV, a mismatch that we already find at the level of LDA. Fixing this inaccuracy hence most likely requires a scheme with full charge self-consistency (in TIER III) or some modification of the bandstructure input. The too strong hybridization with the Oxygen states is also the cause of the strong broadening and inaccurate dispersion of the occupied quasiparticle bands in the $O_p$ models, compared to the experimental data in Ref.~\onlinecite{BackesSrvo3} (top row of Fig. \ref{srvo3_akw_exp}). In this comparison, it appears that the position of the renormalized occupied $t_{2g}$ bands (but not their slope at the Fermi level) is better reproduced by the $t_{2g}$ and $t_{2g}+e_g$ models. 

The $t_{2g}$ dispersions on a longer momentum path (bottom row of Fig. \ref{srvo3_akw_exp}) show similar widths of the occupied $t_{2g}$ bands in all four models, consistent with the local spectra in Fig.~\ref{dosloc_wloc_srvo3}. 

A comparison of the unoccupied parts of the spectral function to Oxygen X-ray absorption spectra (O 1s XAS) from Ref.~\onlinecite{Inoue1994} is shown in the bottom panel of Fig.~\ref{srvo3_pes_xas_dos}. O 1s XAS probes transitions from Oxygen core states to orbitals with partial Oxygen character. Since the Oxygens hybridize with the $t_{2g}$ and $e_g$ orbitals, O 1s XAS allows to probe the unoccupied $t_{2g}$ and $e_g$ states. The XAS spectrum has two notable humps at 0.5 and 2.5 eV, respectively, which are in good agreement with the peaks of the $O_p$ models, which explicitly capture the $O_p$-$t_{2g}$ and $O_p$-$e_{g}$ hybridization. The quasiparticle peak in the $O_p$ models is strongly renormalized in the unoccupied part of the spectrum, compared to the $t_{2g}$ and $e_g$ models, which is consistent with the previously noted stronger renormalization of the Fermi velocity. A comparison to the XAS data further shows that this much stronger renormalization is consistent with experiment.

\begin{figure}[t]
    \centering
    \includegraphics[width=0.4\textwidth]{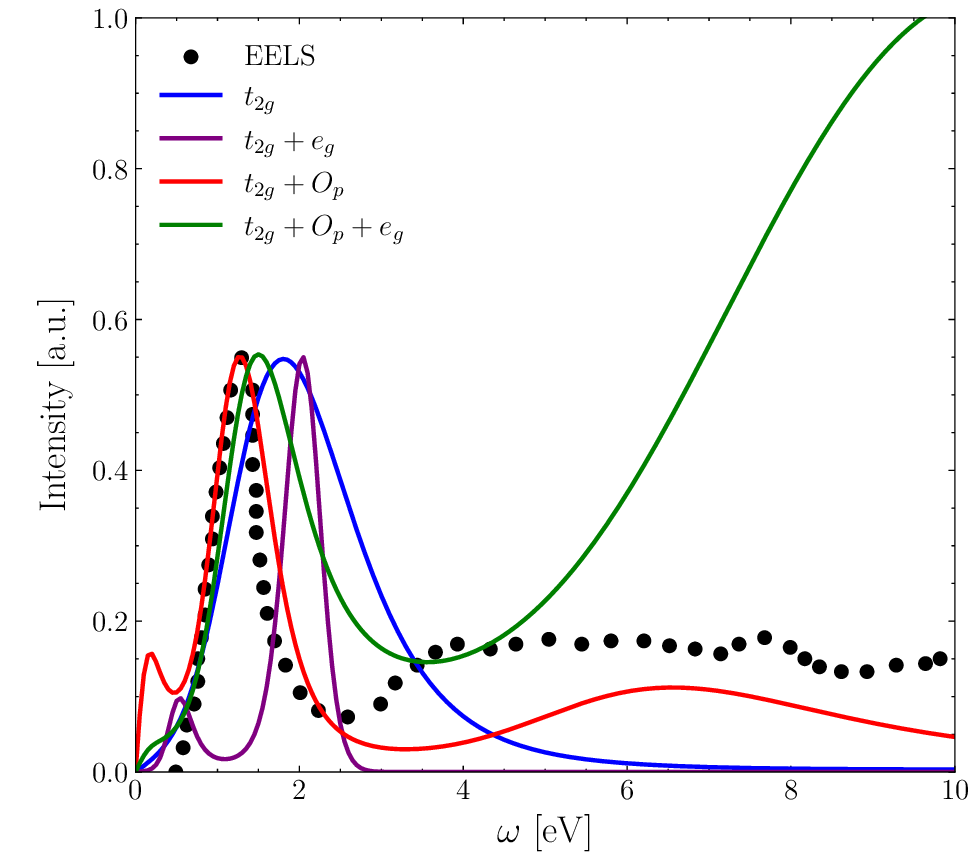}
    \caption{Local loss function $-\mathrm{Im} \sum_q \frac{1}{\epsilon(q,\omega)}$ in different model spaces. Black dots show the experimental values\cite{srvo3_eels_wiley} integrated over the momentum range $0 < q < 4 $ Å$^{-1}$. The calculated curves have been rescaled to match the experimental peak height.}
    \label{srvo3_eels_comp}
\end{figure}

\subsubsection{Plasmon satellites and loss function
\label{k_res_srvo3}}

The nature of the spectral feature at around $3$ eV is still under debate \cite{TomczakSrvo3,PetocchiScreening}. It has been interpreted as an upper Hubbard band \cite{Pavarini2004}, or as a (sub)plasmon feature originating from charge excitations within the $3d$ manifold \cite{NilssonGW_EDMFT}. $GW$+EDMFT, which treats correlation and screening effects in a self-consistent manner, should be a useful tool for distinguishing between these two scenarios. Assuming that the satellite in the occupied part of the spectrum is the lower Hubbard band, the tentative upper Hubbard band position can be estimated from the static limit of the effective local interaction. The Mott gap in the atomic limit is  $\Delta_{\mathrm{at}}:= E_{N+1}-2 E_N + E_{N-1}$, where $N$ is the electron population of the $t_{2g}$ manifold and the lowest energies $E_N$ in the different charge sectors are obtained via Exact Diagonalization (ED) of a Kanamori-type Hamiltonian for the $t_{2g}$ orbitals. Comparing the atomic gap values in  Tab.~\ref{srvo3_table} to the separation between the lower and upper satellites in the local DOS (Fig. \ref{dosloc_wloc_srvo3}) shows that the separations are too large/small with respect to the atomic gap values in all models, with the exception of the $t_{2g}+O_p$ model. There $\Delta_\text{at}$ is compatible with the subband splitting, making it difficult in this case to rule out a Hubbard band interpretation. These findings are consistent with the analysis of the $t_{2g}$ and $t_{2g}+e_g$ models in Refs.~\onlinecite{Boehnke2016,PetocchiScreening}.

\begin{figure*}[ht!]
    \centering
    \includegraphics[width=0.98\textwidth]{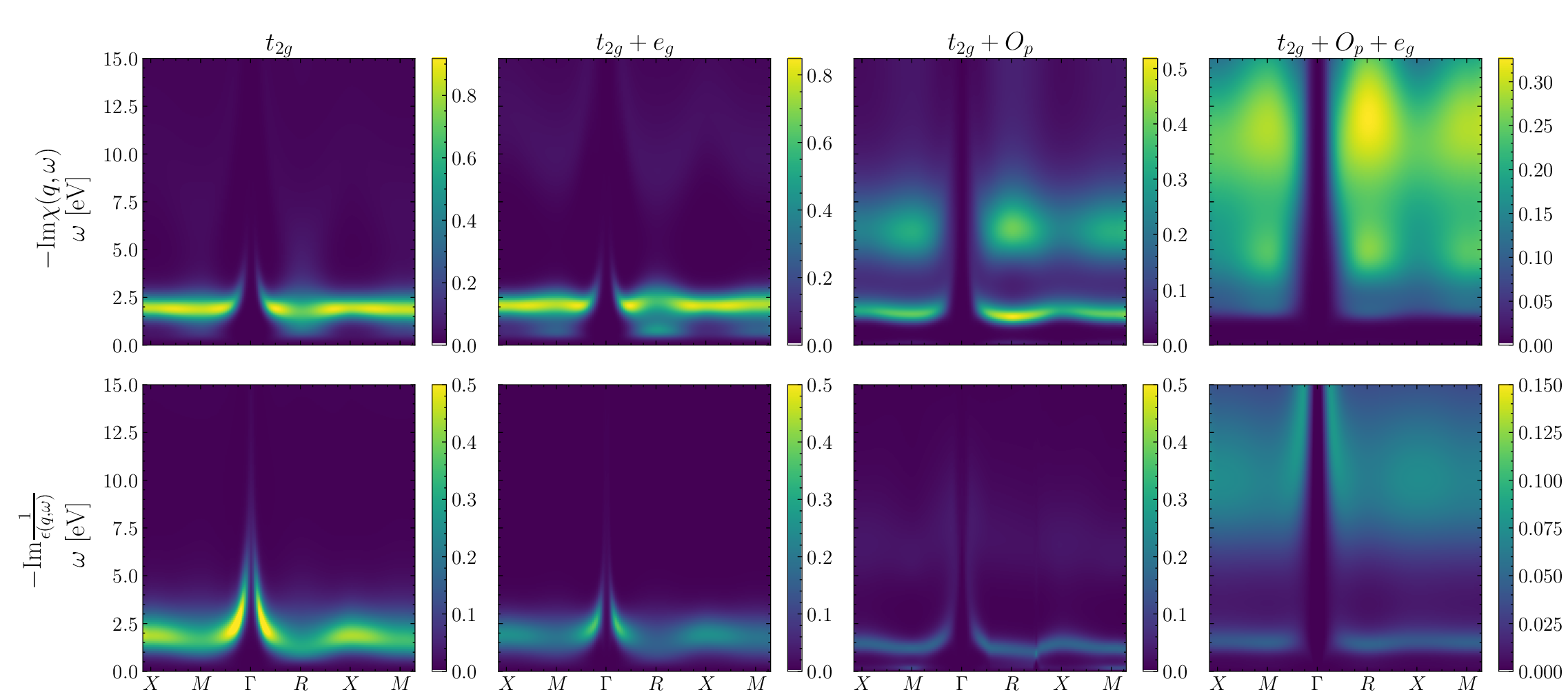}
    \caption{Momentum resolved charge susceptibility (top row) and inverse dielectric function (bottom row) for SrVO$_3$ in the indicated model spaces. The region in the vicinity of the $R$ point for the dielectric function in the $t_{2g}+O_p$ model has been interpolated due to numerical noise.
    }
    \label{chi_eps_inv_srvo3}
\end{figure*}

To test the plasmon scenario as a possible explanation for the high energy spectral feature, we focus on the imaginary part of the local screened interaction $W^{\mathrm{loc}}$, reported in the bottom row of Fig.~\ref{dosloc_wloc_srvo3}, and the loss function $-\mathrm{Im} \sum_q \frac{1}{\epsilon(q,\omega)}$ shown in Fig.~\ref{srvo3_eels_comp}. $W^{\mathrm{loc}}$ features two low energy peaks between 0 and 6 eV in all model spaces. For the $t_{2g}$ and $t_{2g}+e_g$ models, the calculated loss function does not feature the peak above 3 eV, present in the screened interaction, suggesting that the peak in $W^{\mathrm{loc}}$ for these models should originate directly from $U_\mathrm{cRPA}$, which implicitly (through the downfolding) contains the Oxygen screening and whose spectrum indeed has a feature in this energy range, as evidenced in the plot of $U^{(11),(11)}_\mathrm{cRPA}$ in Fig.~\ref{dft_bands_crpa}. 

When the Oxygen states are explicitly included, the loss function recovers the higher energy peak, in reasonable agreement with that of $W^{\mathrm{loc}}$ for the $t_{2g}+O_p$ model, implying that the upper lying peak is mostly due to Oxygen states. This is also evidenced by its dominant oxygen character (Fig.~\ref{dosloc_wloc_srvo3}). The large upper peak in the $t_{2g}+O_p+e_g$ loss function can be traced back to the large $O_p$-$e_g$ pole in $U^{\mathrm{cRPA}}$ in the 3-, 5- and 12-band models (bottom left of Fig. \ref{dft_bands_crpa}), which is now explicitly included in the screening channels of the 14-band model, captured by the loss function. Its relative amplitude is severely overestimated when compared to experiment, potentially due to a lower maximum Matsubara frequency cutoff used in the 14-band model.

The lower lying experimental peak at $\sim 1.5$ eV is captured rather well by all models and is mostly of V $d$ character, with some contribution from the O $p$ states.

A plasmonic excitation should manifest itself as a faint replica of the quasiparticle bandstructure, separated by the plasmon frequency (pole of the local screened interaction). A plasmon feature in the local fermionic spectral function would then be expected at $\omega_\text{qp}+ \omega_\text{pl}$, where $\omega_\text{qp}$ is the position of the quasiparticle peak and $\omega_\text{pl}$ the energy of one of the poles in $\text{Im}W^{\mathrm{loc}}$. Such satellite features are indeed evident in the momentum resolved spectra shown in Fig.~\ref{srvo3_akw_exp}, which resemble the $GW$ + cumulant expansion results in Ref.~\onlinecite{Nakamura2016}. 

More quantitatively, taking the lowest energy poles of $W^{\mathrm{loc}}$  at $1.7,1.4,1.7$ and $1.4$ eV for the $t_{2g}$,  $t_{2g}+e_g$, $t_{2g}+O_p$ and $t_{2g}+O_p+e_g$ models, respectively, 
$\omega_{\text{qp}}+
\omega_{\text{pl}}$ is roughly at $\sim 2.7$, $2.5$, $2.1$ and $2.8$ eV. These values are compatible with the energy region of the right-most hump in the fermionic DOS (top row of Fig. \ref{dosloc_wloc_srvo3}). 

There is thus substantial evidence for interpreting the satellite in the unoccupied spectrum as a (sub-)plasmonic excitation. The dispersion of the plasmon is depicted in Fig.~\ref{chi_eps_inv_srvo3}, which plots the charge susceptibility alongside the momentum-resolved loss function, which should diverge for plasmonic excitations. The negative curvature around the $\Gamma$ point can be likely attributed to band-structure effects similar to those occurring in transition metal dichalcogenides \cite{plasmons_prl,pines_nature}.

The orbital make-up of the low-energy plasmonic excitation is mostly composed of $t_{2g}$-$t_{2g}$ and $t_{2g}$-$e_g$ excitations in all models (see the orbital-resolved plots of $W_{\mathrm{loc}}$ in Fig. \ref{dosloc_wloc_srvo3}), although excitations involving $O_p$ states also remain competitive, likely due to $t_{2g}$-$O_p$ and $e_g$-$O_p$ hybridization.

\begin{figure*}[ht]
    \centering
    \includegraphics[width=0.9\textwidth]{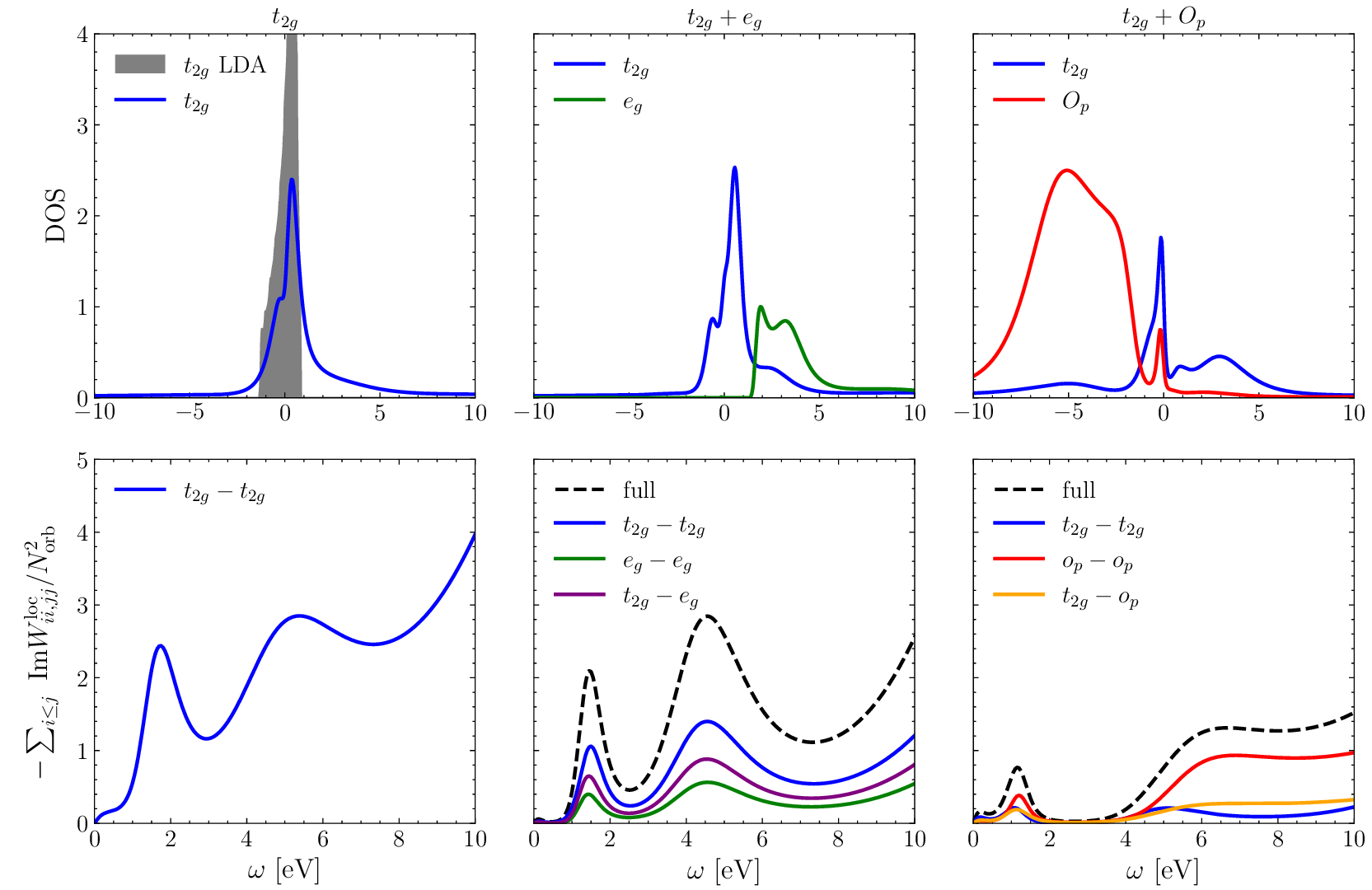}
    \caption{Local spectral functions of SrCrO$_3$ computed within different self-consistent model spaces. The top panels show the fermionic spectra and the bottom panels the bosonic spectra (screened interaction). In the case of the $t_{2g}+e_g$ and $t_{2g}+O_p$ models, the contributions from different subsets of density-density matrix elements to $\text{Im}W^\text{loc}$ are shown by solid lines. 
    }
    \label{dosloc_wloc_srcro3}
\end{figure*}

\subsection{SrCrO$_3$}

SrCrO$_3$ is a less-studied compound,  whose ground state is still under debate due to a difficult high-pressure synthesis procedure. The first experimental measurements \cite{chamberland} suggested that SrCrO$_3$ is a paramagnetic metal, while more recent results
\cite{srcro3_zhou} reported a  paramagnetic insulating behavior. EELS measurements performed
in Ref.~\onlinecite{srcro3_eels}
indicate a paramagnetic metal. These experimental results contradict the numerical DFT calculations
in Ref.~\onlinecite{srcro3_qian}, which predicted a rare antiferromagnetic (AFM) weakly correlated metallic phase, which was later supported by X-ray and O 1s XAS data \cite{zhang_srcro3}. A recent DFT+DMFT investigation \cite{carta2023emergence} found that SrCrO$_3$ can potentially host a charge disproportionated meta-stable insulating state, although implementing charge self-consistency yielded a metallic solution. It is thus interesting to apply the $GW$+EDMFT machinery to this compound.

\begin{table}[b]
    \centering
        \caption{$t_{2g}$ filling, quasiparticle weight, atomic gap and static  interactions for SrCrO$_3$ in the three model spaces.}
    \label{srcro3_table}
    \begin{tabular}{ |p{3cm}|p{1.66cm}|p{1.66cm}|p{1.66cm}| }
    \hline
      & $t_{2g}$ & $t_{2g}+e_g$ & $t_{2g}+O_p$ \\
     \hline
     \hline
    $N_{t_{2g}}$ & 2.00 & 1.83 & 2.95\\
     \hline
     $Z_{t_{2g}}$ (from $\Sigma^\text{EDMFT}$)& 0.50   & 0.54 & 0.16\\
     \hline
     $\Delta_{\mathrm{at}} \ [\mathrm{eV}]$ & 1.4   & 1.1 & 5.8 \\
     \hline
    $\mathcal{U}^{\mathrm{static}}_{t_{2g}}  \ [\mathrm{eV}] $ & 2.3  & 2.3 & 4.6 \\
     \hline
     $\mathcal{J}^{\mathrm{static}}_{t_{2g}}  \ [\mathrm{eV}] $ & 0.3   & 0.4 & 0.6 \\
     \hline
     $U^{\mathrm{cRPA}}_{t_{2g}}  \ [\mathrm{eV}] $ & 3.0   & 3.2 & 6.1 \\
     \hline
    \end{tabular} 
\end{table}

At the LDA level, there are 2 electrons in the $t_{2g}$ derived bands. $GW$+EDMFT yields similar occupations for the $t_{2g}$ and $t_{2g}+e_{g}$ models, while the $t_{2g}$-like orbitals are almost half-filled in the $t_{2g}+O_p$ model, see Tab.~\ref{srcro3_table}. The static $\mathcal{U}_{t_{2g}}$ values are smaller than in the case of SrVO$_3$, but since the filling is larger, the correlation effects are stronger, as indicated by the smaller quasi-particle weight. 

\begin{figure}[t]
    \centering
    \includegraphics[width=0.45\textwidth]{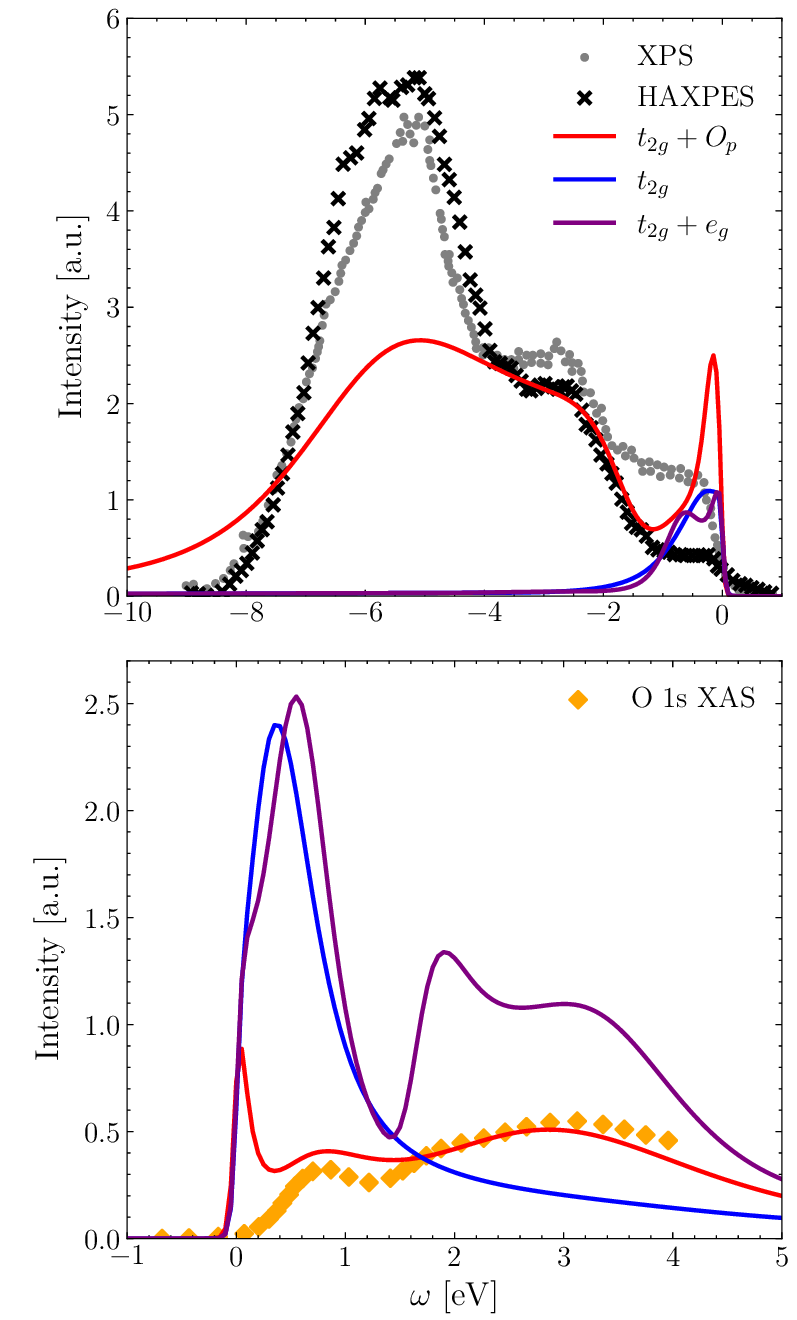}
    \caption{Comparison of the local spectral functions to the experimental XPS/HAXPES/XAS spectra extracted from Ref.~\onlinecite{zhang_srcro3}. The spectral functions have been multiplied by the Fermi-Dirac function $f(\omega)$ and by $1-f(\omega)$ (assuming the temperature $T=298$K) in the upper and lower panels, respectively.}
    \label{srcro3_pes_xas_dos}
\end{figure}

The local fermionic spectral functions are plotted for the three models in the top row of Fig.~\ref{dosloc_wloc_srcro3}. All the solutions are metallic (we restrict our calculation here to the paramagnetic phase). Again, due to stronger bare interactions and larger filling, the renormalization of the $t_{2g}$ bands is more pronounced in the $t_{2g}+O_p$ model than in the other two models. 

The local $t_{2g}$ spectra show satellites below and above the quasi-particle peak, but in the case of the 3- and 5-band models, a comparison between the energy splitting and the atomic gap $\Delta_\text{at}$ (Tab.~\ref{srcro3_table}) shows that these cannot be interpreted as Hubbard bands. The corresponding local screened interaction is plotted in the bottom  row of Fig.~\ref{dosloc_wloc_srcro3}. The lowest energy pole in $\mathrm{Im} W^{\mathrm{loc}}$ at energy $\omega_\text{pl}=1.5-1.7$ eV is compatible with the satellite feature in the fermionic spectra around $\sim 2$ eV, and appears to contribute also to the very weak satellite on the occupied side of the spectrum, consistent with the plasmon interpretation.  The second lowest pole at energy $\omega_\text{pl}\approx 5$ eV may furthermore contribute to the broad tails of the satellites, although in the $t_{2g}+O_p$ case, these contributions are masked by the hybridization with the Oxygen bands (occupied part), and by a possible upper Hubbard band feature (unoccupied part).

\begin{figure*}[ht!]
    \centering
\includegraphics[width=0.9\textwidth]{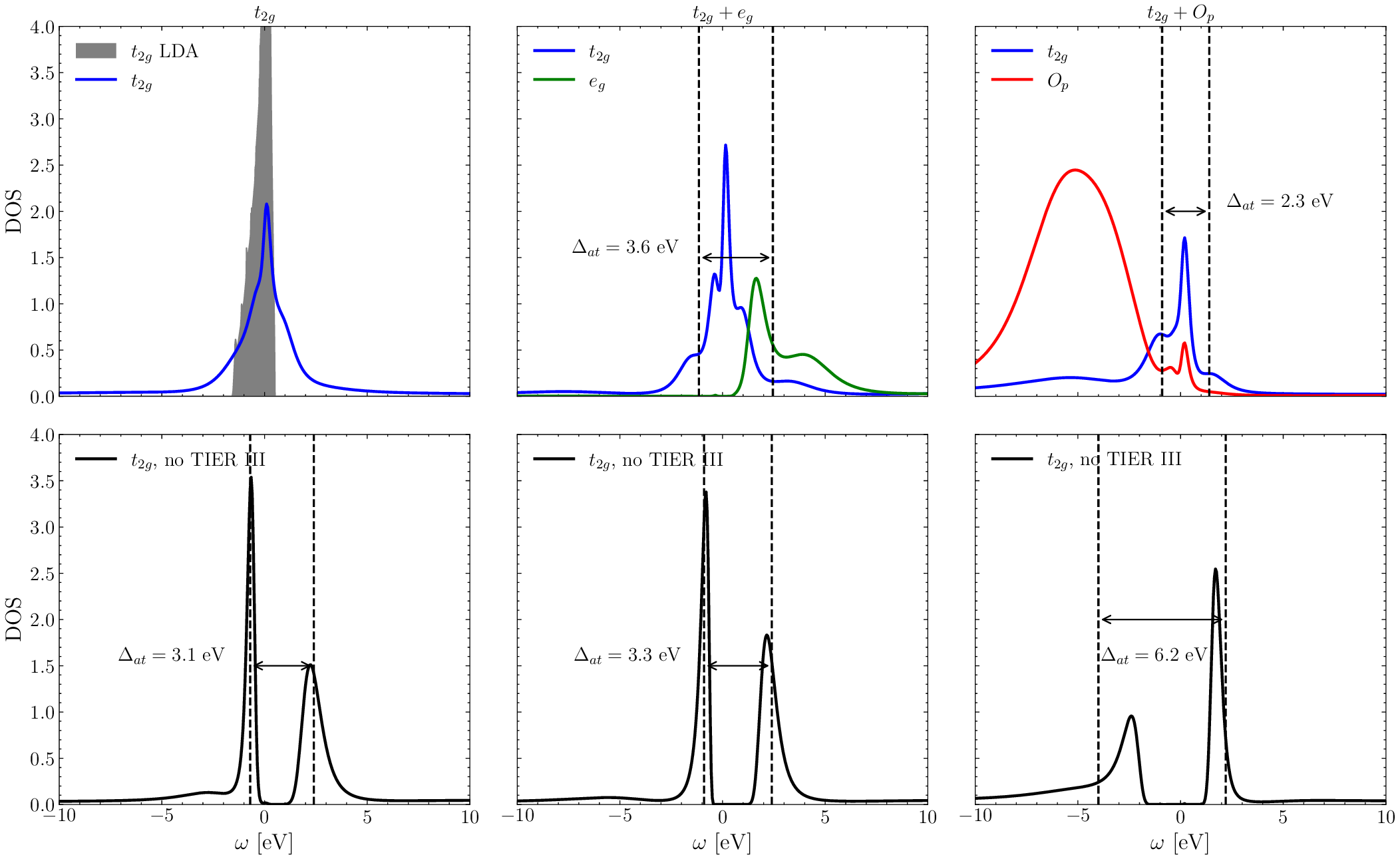}
    \caption{Local fermionic spectral functions of SrMnO$_3$ in the paramagnetic state, computed within the indicated self-consistent model spaces with and without TIER III self-energy contribution. The atomic gap values from Tab.~\ref{srmno3_table} are indicated by the arrows.
    }
    \label{dosloc_srmno3}
\end{figure*}

Figure~\ref{srcro3_pes_xas_dos} shows a comparison of the occupied and unoccupied spectral functions to the  experimental spectra from Ref.~\onlinecite{zhang_srcro3} in the top and bottom panels, respectively. The occupied part of the spectrum has been probed with X-ray photoemission spectroscopy (XPS/HAXPES) \cite{zhang_srcro3} and the unoccupied part with O 1s XAS. We note a very good agreement for the $t_{2g}+O_p$ model, which is able to reproduce all features in the occupied Oxygen and in the unoccupied $t_{2g}+O_p$ spectrum. Only the presence of a quasi-particle peak is not confirmed by the experiments, but as mentioned above, the experimental situation concerning the metallic or insulating behavior is not yet settled.  Nevertheless, we note that for all models, the width of the occupied quasi-particle peak is comparable to the extent of the right-most step feature between $-1$ and 0 eV in the experimental spectra, which supports the theoretically predicted metallic solution. The first peak near 0.8 eV in the O 1s XAS spectra could be associated either with the first satellite in the $t_{2g}+O_p$ model spectrum, or with the much broader quasi-particle peak in the $t_{2g}$ and $t_{2g}+e_g$ spectra. 
Similarly, the second broader satellite near 3 eV, which perfectly agrees in position and shape with the $t_{2g}+O_p$ spectrum, could originate also from $e_g$ states, as seen from the comparison to the corresponding hump in the $t_{2g}+e_g$ spectrum. A signal from the $e_g$ states is expected in the O 1s XAS spectra, since the Oxygen states also hybridize with the $e_g$ orbitals.

\subsection{SrMnO$_3$}

\begin{table}[b]
    \centering
    \begin{tabular}{ |p{2cm}|p{2cm}|p{2cm}|p{2cm}| }
    \hline
      & $t_{2g}$ & $t_{2g}+e_g$ & $t_{2g}+O_p$ \\
     \hline
     \hline
    $N_{t_{2g}}$ & 3.00 (3.00) & 2.79 (3.00) & 3.97 (3.16)\\
     \hline
     $Z_{t_{2g}}$ & 0.36   & 0.43 & 0.18\\
     \hline
     $\Delta_{\mathrm{at}} \ [\mathrm{eV}]$ & 2.7 (3.1) & 3.6 (3.3) & 2.3 (6.2)\\
     \hline
    $\mathcal{U}^{\mathrm{static}}_{t_{2g}}  \ [\mathrm{eV}] $ & 2.1 (2.3) & 2.5 (2.5) & 4.1 (4.8)\\
     \hline
     $\mathcal{J}^{\mathrm{static}}_{t_{2g}}  \ [\mathrm{eV}] $ & 0.3 (0.4)  & 0.3 (0.4) & 0.6 (0.7)\\
     \hline
     $U^{\mathrm{cRPA}}_{t_{2g}}  \ [\mathrm{eV}] $ & 2.3   & 2.6 & 5.0 \\
     \hline
    \end{tabular} 
    \caption{$t_{2g}$ filling, quasiparticle weight, atomic gap and static interactions for SrMnO$_3$ in the three model spaces. The values in brackets are for a calculation without TIER III self-energy contribution, which yields an insulating solution.}
    \label{srmno3_table}
\end{table}

SrMnO$_3$ is an insulating, antiferromagnetic perovskite compound, with a N\'eel temperature $T_{\mathrm{N\acute{e}el}}$ between 233~K \cite{chmaissem_srmno3} and 266~K \cite{takeda_srmno3}.

Its Manganese cation hosts 3 $d$-electrons in the unit cell, which at the LDA level occupy almost entirely the $t_{2g}$ bands (Fig.~\ref{dft_bands_crpa}). In paramagnetic $GW$+EDMFT calculations at $T > T_{\mathrm{N\acute{e}el}}$, the local spectral function exhibits metallic behavior, with a quasi-particle peak in the vicinity of the Fermi level (Fig. \ref{dosloc_srmno3}), for all the low-energy subspaces considered. Upon removing the TIER III ($G^0W^0$) self-energy contribution (which contains metallic screening from the DFT bandstructure), however, one recovers an electronic gap (bottom row of Fig. \ref{dosloc_srmno3}). This suggests a sensitivity of the method to the DFT starting point, since the metallicity of the DFT solution affects the $GW+$EDMFT result even after self-consistency.

While the gap is missing in the full calculation, we may nevertheless proceed with the analysis of the higher-energy structures in the local spectral function. Here, we note the appearance of satellites in the $t_{2g}+e_g$ and $t_{2g}+O_p$ model spectra, whose separation is in good agreement with the gap $\Delta_\text{at}$ computed in the atomic limit via ED, see Tab.~\ref{srmno3_table} and the arrows in Fig.~\ref{dosloc_srmno3}. We thus interpret these features as Hubbard bands. The Hubbard satellites cannot be resolved in the $t_{2g}$ model, but they become  prominent once we add the $e_g$ states. This shows that the $e_g$ states have a significant effect on the low-energy screening in SrMnO$_3$, and hence on the correlation strength, as confirmed by the static impurity interaction values in Tab.~\ref{srmno3_table}. A more active role of the $e_g$ states, compared to SrVO$_3$ and SrCrO$_3$,  is already expected from the LDA bandstructure (Fig.~\ref{dft_bands_crpa}), where the $e_g$ bands for SrMnO$_3$ overlap with the $t_{2g}$ bands. 

Incorporating the Oxygen orbitals into the SC space again results in a jump in the filling of the $t_{2g}$ orbitals, as in the case of SrVO$_3$: the $t_{2g}$ shell now has nominally one extra electron coming from the hybridization with the Oxygens (see the prominent hump in Fig.~\ref{dosloc_srmno3} at $\omega\approx -5$~eV). The quasi-particle weight, estimated from $\Sigma^\text{EDMFT}$ undergoes a substantial renormalization by a factor of two, compared to the $t_{2g}$ model, which is more pronounced than what we observed for SrVO$_3$.

\section{\label{disc} Discussion}

We have systematically tested the multi-tier $GW$+EDMFT framework on the perovskite compounds SrXO$_3$ (X=V, Cr, Mn) by comparing the low-energy electronic structures calculated in different energy windows. If the downfolding to these model subspaces, as well as the solutions of the corresponding many-body problems would be implemented with exact methods, the results would be consistent (up to hybridization effects).

For practical reasons the downfolding has been implemented here via cRPA and single-shot $G_0W_0$, while TIER II is treated at the $GW$ level (with local vertex corrections), and TIER III at the EDMFT level (with density-density interactions). It is thus interesting to test to what extent this {\it ab-initio} framework produces consistent results for different model spaces. While inaccuracies may be expected for very small low-energy subspaces, one would expect systematic improvements as the self-consistently solved model spaces are enlarged.

To judge if this is indeed the case, we compared the calculated results to available PES and XAS data. For SrVO$_3$, $GW$+EDMFT predicts a moderately correlated metallic phase in all model spaces, with the $O_p$ models yielding mass enhancements closest to experiment. The reduced band velocity near the Fermi level is a consequence of a strong renormalization of the bands in the unoccupied part of the spectrum, while in the occupied part, the $t_{2g}$ and $t_{2g}+e_g$ models actually produce a better agreement with the PES measurements along the $\Gamma$-X path. The much stronger renormalization of the unoccupied bands in the $O_p$ models is consistent with the O 1s XAS spectrum, which due to hybridization between the $O_p$ and $t_{2g}/e_g$ orbitals provides information on the unoccupied $d$ states. The much discussed satellite at $-1.5$~eV in the local spectra (which has been interpreted as a lower Hubbard band or as a signal from Oxygen vacancies) is weakly present as a shoulder feature at -1 eV in the $O_p$ models, while the satellite around 3~eV  could be interpreted as a plasmon. The position of the Oxygen bands in the $t_{2g}+O_p$ calculations is about 1~eV too high, compared to the PES data, but this is already the case in the LDA input. Our calculations are not self-consistent in TIER III, and a charge-self consistent implementation could potentially correct this inaccuracy.       

In the case of SrCrO$_3$ we have demonstrated an excellent agreement between the $t_{2g}+O_p$ model results and the available XPS and O 1s XAS data for the Oxygen and unoccupied $t_{2g}$ states, while the signature of the quasi-particle peak is not clearly seen in XPS. The low-energy peak in the O 1s XAS would however be equally consistent with the (much wider) quasi-particle band of the $t_{2g}$ and  $t_{2g}+e_g$ models, so that it is not clear in this case if the stronger correlation effects in the $t_{2g}+O_p$ description improve the agreement with experiment. Again, the high-energy satellite near $3$~eV overlaps with the $e_g$ states, so that the XAS signal presumably detects not only the plasmon satellite of the $t_{2g}$ bands, but also the $e_g$ states. 

For SrMnO$_3$, our paramagnetic calculations in all three model subspaces predict a metallic system, likely caused by the metallicity of the DFT starting point.

Our $t_{2g}+e_g$ and $t_{2g}+O_p$ simulations produced Hubbard bands in the (metallic) spectra, in contrast to the $t_{2g}$-only calculations. This shows that in this system, the $e_g$ states (which are very close to the Fermi level) play a relevant role in the low-energy screening of the $t_{2g}$ electrons \cite{PetocchiScreening}. 

Overall, our calculations demonstrated a fair agreement between the first-principles results for the different model subspaces, and between the calculated and experimentally measured data, bearing in mind the absence of adjustable parameters in $GW$+EDMFT. Our study, however, also showed that there exist significant discrepancies between the results from the $t_{2g}(+e_g)$  and $O_p$ model spaces, with the latter exhibiting stronger correlation effects. Since the $t_{2g}+O_p$ and $t_{2g}+e_g+O_p$ calculations appear to be in overall better agreement with the experiments, this hints at an overestimation of screening in the cRPA downfolding. An accurate treatment of screening from low-energy bands may require more sophisticated downfolding schemes \cite{Shinaoka2015,Honerkamp2018}. 

\section{\label{method} Methods}

The DFT calculations were performed with the all-electron full-potential linearized augmented plane-wave (FLAPW) code FLEUR \cite{fleur} on a $20 \times 20 \times 20$ {\bf k}-point mesh, and the cRPA and $G^0W^0$ downfolding with a customized SPEX code \cite{spex}. Wannier functions from the Wannier90 library \cite{wannier90} were used to define the models. The self-consistent multi-tier $GW$+EDMFT calculations were performed with the implementation described in Refs.~\onlinecite{NilssonGW_EDMFT,Boehnke2016}, using a
CT-HYB quantum impurity solver for multi-orbital models with retarded interactions \cite{ct_hyb,ct_hyb_retarded}. A detailed description can be found in the Supplementary Material. A $10 \times 10 \times 10$ {\bf k}-point grid and an inverse temperature of $\beta=10 \text{ eV}^{-1}$ were used in all self-consistent $GW$-EDMFT calculations, with 477 Matsubara frequencies for the 3-, 5- and 12-band models and 238 for the 14-band model. The charge susceptibilities and dielectric functions were computed with the procedures described in the Supplementary Material.

\vspace{10mm}
\begin{acknowledgments}
This work was supported by the Swiss National Science Foundation via NCCR Marvel and SNSF Grant No.~200021-196966. The calculations were run on the beo05 cluster at the University of Fribourg. F.P. acknowledges helpful discussions with Antoine Georges.
\end{acknowledgments}

\bibliography{srxo3_npj}

\begin{thebibliography}{41}%
\makeatletter
\providecommand \@ifxundefined [1]{%
 \@ifx{#1\undefined}
}%
\providecommand \@ifnum [1]{%
 \ifnum #1\expandafter \@firstoftwo
 \else \expandafter \@secondoftwo
 \fi
}%
\providecommand \@ifx [1]{%
 \ifx #1\expandafter \@firstoftwo
 \else \expandafter \@secondoftwo
 \fi
}%
\providecommand \natexlab [1]{#1}%
\providecommand \enquote  [1]{``#1''}%
\providecommand \bibnamefont  [1]{#1}%
\providecommand \bibfnamefont [1]{#1}%
\providecommand \citenamefont [1]{#1}%
\providecommand \href@noop [0]{\@secondoftwo}%
\providecommand \href [0]{\begingroup \@sanitize@url \@href}%
\providecommand \@href[1]{\@@startlink{#1}\@@href}%
\providecommand \@@href[1]{\endgroup#1\@@endlink}%
\providecommand \@sanitize@url [0]{\catcode `\\12\catcode `\$12\catcode `\&12\catcode `\#12\catcode `\^12\catcode `\_12\catcode `\%12\relax}%
\providecommand \@@startlink[1]{}%
\providecommand \@@endlink[0]{}%
\providecommand \url  [0]{\begingroup\@sanitize@url \@url }%
\providecommand \@url [1]{\endgroup\@href {#1}{\urlprefix }}%
\providecommand \urlprefix  [0]{URL }%
\providecommand \Eprint [0]{\href }%
\providecommand \doibase [0]{https://doi.org/}%
\providecommand \selectlanguage [0]{\@gobble}%
\providecommand \bibinfo  [0]{\@secondoftwo}%
\providecommand \bibfield  [0]{\@secondoftwo}%
\providecommand \translation [1]{[#1]}%
\providecommand \BibitemOpen [0]{}%
\providecommand \bibitemStop [0]{}%
\providecommand \bibitemNoStop [0]{.\EOS\space}%
\providecommand \EOS [0]{\spacefactor3000\relax}%
\providecommand \BibitemShut  [1]{\csname bibitem#1\endcsname}%
\let\auto@bib@innerbib\@empty
\bibitem [{\citenamefont {Dagotto}(2005)}]{Dagotto2005}%
  \BibitemOpen
  \bibfield  {author} {\bibinfo {author} {\bibfnamefont {E.}~\bibnamefont {Dagotto}},\ }\bibfield  {title} {\bibinfo {title} {Complexity in strongly correlated electronic systems},\ }\href {https://doi.org/10.1126/science.1107559} {\bibfield  {journal} {\bibinfo  {journal} {Science}\ }\textbf {\bibinfo {volume} {309}},\ \bibinfo {pages} {257} (\bibinfo {year} {2005})},\ \Eprint {https://arxiv.org/abs/https://www.science.org/doi/pdf/10.1126/science.1107559} {https://www.science.org/doi/pdf/10.1126/science.1107559} \BibitemShut {NoStop}%
\bibitem [{\citenamefont {Hohenberg}\ and\ \citenamefont {Kohn}(1964)}]{Hohenberg1964}%
  \BibitemOpen
  \bibfield  {author} {\bibinfo {author} {\bibfnamefont {P.}~\bibnamefont {Hohenberg}}\ and\ \bibinfo {author} {\bibfnamefont {W.}~\bibnamefont {Kohn}},\ }\bibfield  {title} {\bibinfo {title} {Inhomogeneous electron gas},\ }\href {https://doi.org/10.1103/PhysRev.136.B864} {\bibfield  {journal} {\bibinfo  {journal} {Phys. Rev.}\ }\textbf {\bibinfo {volume} {136}},\ \bibinfo {pages} {B864} (\bibinfo {year} {1964})}\BibitemShut {NoStop}%
\bibitem [{\citenamefont {Kohn}\ and\ \citenamefont {Sham}(1965)}]{Kohn1965}%
  \BibitemOpen
  \bibfield  {author} {\bibinfo {author} {\bibfnamefont {W.}~\bibnamefont {Kohn}}\ and\ \bibinfo {author} {\bibfnamefont {L.~J.}\ \bibnamefont {Sham}},\ }\bibfield  {title} {\bibinfo {title} {Self-consistent equations including exchange and correlation effects},\ }\href {https://doi.org/10.1103/PhysRev.140.A1133} {\bibfield  {journal} {\bibinfo  {journal} {Phys. Rev.}\ }\textbf {\bibinfo {volume} {140}},\ \bibinfo {pages} {A1133} (\bibinfo {year} {1965})}\BibitemShut {NoStop}%
\bibitem [{\citenamefont {Georges}\ \emph {et~al.}(1996)\citenamefont {Georges}, \citenamefont {Kotliar}, \citenamefont {Krauth},\ and\ \citenamefont {Rozenberg}}]{Georges1996}%
  \BibitemOpen
  \bibfield  {author} {\bibinfo {author} {\bibfnamefont {A.}~\bibnamefont {Georges}}, \bibinfo {author} {\bibfnamefont {G.}~\bibnamefont {Kotliar}}, \bibinfo {author} {\bibfnamefont {W.}~\bibnamefont {Krauth}},\ and\ \bibinfo {author} {\bibfnamefont {M.~J.}\ \bibnamefont {Rozenberg}},\ }\bibfield  {title} {\bibinfo {title} {Dynamical mean-field theory of strongly correlated fermion systems and the limit of infinite dimensions},\ }\href {https://doi.org/10.1103/RevModPhys.68.13} {\bibfield  {journal} {\bibinfo  {journal} {Rev. Mod. Phys.}\ }\textbf {\bibinfo {volume} {68}},\ \bibinfo {pages} {13} (\bibinfo {year} {1996})}\BibitemShut {NoStop}%
\bibitem [{\citenamefont {Haule}\ and\ \citenamefont {Birol}(2015)}]{HauleDFT_DMFT}%
  \BibitemOpen
  \bibfield  {author} {\bibinfo {author} {\bibfnamefont {K.}~\bibnamefont {Haule}}\ and\ \bibinfo {author} {\bibfnamefont {T.}~\bibnamefont {Birol}},\ }\bibfield  {title} {\bibinfo {title} {Free energy from stationary implementation of the $\mathrm{DFT}+\mathrm{DMFT}$ functional},\ }\href {https://doi.org/10.1103/PhysRevLett.115.256402} {\bibfield  {journal} {\bibinfo  {journal} {Phys. Rev. Lett.}\ }\textbf {\bibinfo {volume} {115}},\ \bibinfo {pages} {256402} (\bibinfo {year} {2015})}\BibitemShut {NoStop}%
\bibitem [{\citenamefont {Shim}\ \emph {et~al.}(2007)\citenamefont {Shim}, \citenamefont {Haule},\ and\ \citenamefont {Kotliar}}]{haule_science}%
  \BibitemOpen
  \bibfield  {author} {\bibinfo {author} {\bibfnamefont {J.~H.}\ \bibnamefont {Shim}}, \bibinfo {author} {\bibfnamefont {K.}~\bibnamefont {Haule}},\ and\ \bibinfo {author} {\bibfnamefont {G.}~\bibnamefont {Kotliar}},\ }\bibfield  {title} {\bibinfo {title} {Modeling the localized-to-itinerant electronic transition in the heavy fermion system $\mathrm{CeIrIn}_5$},\ }\href {https://doi.org/10.1126/science.1149064} {\bibfield  {journal} {\bibinfo  {journal} {Science}\ }\textbf {\bibinfo {volume} {318}},\ \bibinfo {pages} {1615} (\bibinfo {year} {2007})}\BibitemShut {NoStop}%
\bibitem [{\citenamefont {Karolak}\ \emph {et~al.}(2010)\citenamefont {Karolak}, \citenamefont {Ulm}, \citenamefont {Wehling}, \citenamefont {Mazurenko}, \citenamefont {Poteryaev},\ and\ \citenamefont {Lichtenstein}}]{KarolakDC_DFT}%
  \BibitemOpen
  \bibfield  {author} {\bibinfo {author} {\bibfnamefont {M.}~\bibnamefont {Karolak}}, \bibinfo {author} {\bibfnamefont {G.}~\bibnamefont {Ulm}}, \bibinfo {author} {\bibfnamefont {T.}~\bibnamefont {Wehling}}, \bibinfo {author} {\bibfnamefont {V.}~\bibnamefont {Mazurenko}}, \bibinfo {author} {\bibfnamefont {A.}~\bibnamefont {Poteryaev}},\ and\ \bibinfo {author} {\bibfnamefont {A.}~\bibnamefont {Lichtenstein}},\ }\bibfield  {title} {\bibinfo {title} {Double counting in lda+dmft---the example of nio},\ }\href {https://doi.org/https://doi.org/10.1016/j.elspec.2010.05.021} {\bibfield  {journal} {\bibinfo  {journal} {Journal of Electron Spectroscopy and Related Phenomena}\ }\textbf {\bibinfo {volume} {181}},\ \bibinfo {pages} {11} (\bibinfo {year} {2010})},\ \bibinfo {note} {proceedings of International Workshop on Strong Correlations and Angle-Resolved Photoemission Spectroscopy 2009}\BibitemShut {NoStop}%
\bibitem [{\citenamefont {Biermann}\ \emph {et~al.}(2003)\citenamefont {Biermann}, \citenamefont {Aryasetiawan},\ and\ \citenamefont {Georges}}]{Biermann2003}%
  \BibitemOpen
  \bibfield  {author} {\bibinfo {author} {\bibfnamefont {S.}~\bibnamefont {Biermann}}, \bibinfo {author} {\bibfnamefont {F.}~\bibnamefont {Aryasetiawan}},\ and\ \bibinfo {author} {\bibfnamefont {A.}~\bibnamefont {Georges}},\ }\bibfield  {title} {\bibinfo {title} {First-principles approach to the electronic structure of strongly correlated systems: Combining the $gw$ approximation and dynamical mean-field theory},\ }\href {https://doi.org/10.1103/PhysRevLett.90.086402} {\bibfield  {journal} {\bibinfo  {journal} {Phys. Rev. Lett.}\ }\textbf {\bibinfo {volume} {90}},\ \bibinfo {pages} {086402} (\bibinfo {year} {2003})}\BibitemShut {NoStop}%
\bibitem [{\citenamefont {Boehnke}\ \emph {et~al.}(2016)\citenamefont {Boehnke}, \citenamefont {Nilsson}, \citenamefont {Aryasetiawan},\ and\ \citenamefont {Werner}}]{Boehnke2016}%
  \BibitemOpen
  \bibfield  {author} {\bibinfo {author} {\bibfnamefont {L.}~\bibnamefont {Boehnke}}, \bibinfo {author} {\bibfnamefont {F.}~\bibnamefont {Nilsson}}, \bibinfo {author} {\bibfnamefont {F.}~\bibnamefont {Aryasetiawan}},\ and\ \bibinfo {author} {\bibfnamefont {P.}~\bibnamefont {Werner}},\ }\bibfield  {title} {\bibinfo {title} {When strong correlations become weak: Consistent merging of $gw$ and dmft},\ }\href {https://doi.org/10.1103/PhysRevB.94.201106} {\bibfield  {journal} {\bibinfo  {journal} {Phys. Rev. B}\ }\textbf {\bibinfo {volume} {94}},\ \bibinfo {pages} {201106} (\bibinfo {year} {2016})}\BibitemShut {NoStop}%
\bibitem [{\citenamefont {Nilsson}\ \emph {et~al.}(2017)\citenamefont {Nilsson}, \citenamefont {Boehnke}, \citenamefont {Werner},\ and\ \citenamefont {Aryasetiawan}}]{NilssonGW_EDMFT}%
  \BibitemOpen
  \bibfield  {author} {\bibinfo {author} {\bibfnamefont {F.}~\bibnamefont {Nilsson}}, \bibinfo {author} {\bibfnamefont {L.}~\bibnamefont {Boehnke}}, \bibinfo {author} {\bibfnamefont {P.}~\bibnamefont {Werner}},\ and\ \bibinfo {author} {\bibfnamefont {F.}~\bibnamefont {Aryasetiawan}},\ }\bibfield  {title} {\bibinfo {title} {Multitier self-consistent $gw+\text{EDMFT}$},\ }\href {https://doi.org/10.1103/PhysRevMaterials.1.043803} {\bibfield  {journal} {\bibinfo  {journal} {Phys. Rev. Mater.}\ }\textbf {\bibinfo {volume} {1}},\ \bibinfo {pages} {043803} (\bibinfo {year} {2017})}\BibitemShut {NoStop}%
\bibitem [{\citenamefont {Kang}\ \emph {et~al.}(2023)\citenamefont {Kang}, \citenamefont {Semon}, \citenamefont {Melnick}, \citenamefont {Kotliar},\ and\ \citenamefont {Choi}}]{Choi2023}%
  \BibitemOpen
  \bibfield  {author} {\bibinfo {author} {\bibfnamefont {B.}~\bibnamefont {Kang}}, \bibinfo {author} {\bibfnamefont {P.}~\bibnamefont {Semon}}, \bibinfo {author} {\bibfnamefont {C.}~\bibnamefont {Melnick}}, \bibinfo {author} {\bibfnamefont {G.}~\bibnamefont {Kotliar}},\ and\ \bibinfo {author} {\bibfnamefont {S.}~\bibnamefont {Choi}},\ }\bibfield  {title} {\bibinfo {title} {Comdmft v.2.0: Fully self-consistent ab initio gw+edmft for the electronic structure of correlated quantum materials},\ }\href {https://arxiv.org/abs/2310.04613} {\bibfield  {journal} {\bibinfo  {journal} {arXiv:2310.04613}\ } (\bibinfo {year} {2023})}\BibitemShut {NoStop}%
\bibitem [{\citenamefont {Hedin}(1965)}]{HedinGW}%
  \BibitemOpen
  \bibfield  {author} {\bibinfo {author} {\bibfnamefont {L.}~\bibnamefont {Hedin}},\ }\bibfield  {title} {\bibinfo {title} {New method for calculating the one-particle green's function with application to the electron-gas problem},\ }\href {https://doi.org/10.1103/PhysRev.139.A796} {\bibfield  {journal} {\bibinfo  {journal} {Phys. Rev.}\ }\textbf {\bibinfo {volume} {139}},\ \bibinfo {pages} {A796} (\bibinfo {year} {1965})}\BibitemShut {NoStop}%
\bibitem [{\citenamefont {Honerkamp}\ \emph {et~al.}(2018)\citenamefont {Honerkamp}, \citenamefont {Shinaoka}, \citenamefont {Assaad},\ and\ \citenamefont {Werner}}]{Honerkamp2018}%
  \BibitemOpen
  \bibfield  {author} {\bibinfo {author} {\bibfnamefont {C.}~\bibnamefont {Honerkamp}}, \bibinfo {author} {\bibfnamefont {H.}~\bibnamefont {Shinaoka}}, \bibinfo {author} {\bibfnamefont {F.~F.}\ \bibnamefont {Assaad}},\ and\ \bibinfo {author} {\bibfnamefont {P.}~\bibnamefont {Werner}},\ }\bibfield  {title} {\bibinfo {title} {Limitations of constrained random phase approximation downfolding},\ }\href {https://doi.org/10.1103/PhysRevB.98.235151} {\bibfield  {journal} {\bibinfo  {journal} {Phys. Rev. B}\ }\textbf {\bibinfo {volume} {98}},\ \bibinfo {pages} {235151} (\bibinfo {year} {2018})}\BibitemShut {NoStop}%
\bibitem [{\citenamefont {Morikawa}\ \emph {et~al.}(1995)\citenamefont {Morikawa}, \citenamefont {Mizokawa}, \citenamefont {Kobayashi}, \citenamefont {Fujimori}, \citenamefont {Eisaki}, \citenamefont {Uchida}, \citenamefont {Iga},\ and\ \citenamefont {Nishihara}}]{Morikawa1995}%
  \BibitemOpen
  \bibfield  {author} {\bibinfo {author} {\bibfnamefont {K.}~\bibnamefont {Morikawa}}, \bibinfo {author} {\bibfnamefont {T.}~\bibnamefont {Mizokawa}}, \bibinfo {author} {\bibfnamefont {K.}~\bibnamefont {Kobayashi}}, \bibinfo {author} {\bibfnamefont {A.}~\bibnamefont {Fujimori}}, \bibinfo {author} {\bibfnamefont {H.}~\bibnamefont {Eisaki}}, \bibinfo {author} {\bibfnamefont {S.}~\bibnamefont {Uchida}}, \bibinfo {author} {\bibfnamefont {F.}~\bibnamefont {Iga}},\ and\ \bibinfo {author} {\bibfnamefont {Y.}~\bibnamefont {Nishihara}},\ }\bibfield  {title} {\bibinfo {title} {Spectral weight transfer and mass renormalization in mott-hubbard systems ${\mathrm{srvo}}_{3}$ and ${\mathrm{cavo}}_{3}$: Influence of long-range coulomb interaction},\ }\href {https://doi.org/10.1103/PhysRevB.52.13711} {\bibfield  {journal} {\bibinfo  {journal} {Phys. Rev. B}\ }\textbf {\bibinfo {volume} {52}},\ \bibinfo {pages} {13711} (\bibinfo {year} {1995})}\BibitemShut {NoStop}%
\bibitem [{\citenamefont {Sekiyama}\ \emph {et~al.}(2004)\citenamefont {Sekiyama}, \citenamefont {Fujiwara}, \citenamefont {Imada}, \citenamefont {Suga}, \citenamefont {Eisaki}, \citenamefont {Uchida}, \citenamefont {Takegahara}, \citenamefont {Harima}, \citenamefont {Saitoh}, \citenamefont {Nekrasov}, \citenamefont {Keller}, \citenamefont {Kondakov}, \citenamefont {Kozhevnikov}, \citenamefont {Pruschke}, \citenamefont {Held}, \citenamefont {Vollhardt},\ and\ \citenamefont {Anisimov}}]{Sekiyama2004}%
  \BibitemOpen
  \bibfield  {author} {\bibinfo {author} {\bibfnamefont {A.}~\bibnamefont {Sekiyama}}, \bibinfo {author} {\bibfnamefont {H.}~\bibnamefont {Fujiwara}}, \bibinfo {author} {\bibfnamefont {S.}~\bibnamefont {Imada}}, \bibinfo {author} {\bibfnamefont {S.}~\bibnamefont {Suga}}, \bibinfo {author} {\bibfnamefont {H.}~\bibnamefont {Eisaki}}, \bibinfo {author} {\bibfnamefont {S.~I.}\ \bibnamefont {Uchida}}, \bibinfo {author} {\bibfnamefont {K.}~\bibnamefont {Takegahara}}, \bibinfo {author} {\bibfnamefont {H.}~\bibnamefont {Harima}}, \bibinfo {author} {\bibfnamefont {Y.}~\bibnamefont {Saitoh}}, \bibinfo {author} {\bibfnamefont {I.~A.}\ \bibnamefont {Nekrasov}}, \bibinfo {author} {\bibfnamefont {G.}~\bibnamefont {Keller}}, \bibinfo {author} {\bibfnamefont {D.~E.}\ \bibnamefont {Kondakov}}, \bibinfo {author} {\bibfnamefont {A.~V.}\ \bibnamefont {Kozhevnikov}}, \bibinfo {author} {\bibfnamefont {T.}~\bibnamefont {Pruschke}}, \bibinfo {author} {\bibfnamefont {K.}~\bibnamefont {Held}}, \bibinfo {author} {\bibfnamefont {D.}~\bibnamefont {Vollhardt}},\ and\ \bibinfo {author} {\bibfnamefont {V.~I.}\ \bibnamefont {Anisimov}},\ }\bibfield  {title} {\bibinfo {title} {Mutual experimental and theoretical validation of bulk photoemission spectra of ${\mathrm{s}\mathrm{r}}_{1\ensuremath{-}x}{\mathrm{c}\mathrm{a}}_{x}{\mathrm{v}\mathrm{o}}_{3}$},\ }\href {https://doi.org/10.1103/PhysRevLett.93.156402} {\bibfield  {journal} {\bibinfo  {journal} {Phys. Rev. Lett.}\ }\textbf {\bibinfo {volume} {93}},\ \bibinfo {pages} {156402} (\bibinfo {year} {2004})}\BibitemShut {NoStop}%
\bibitem [{\citenamefont {Pavarini}\ \emph {et~al.}(2004)\citenamefont {Pavarini}, \citenamefont {Biermann}, \citenamefont {Poteryaev}, \citenamefont {Lichtenstein}, \citenamefont {Georges},\ and\ \citenamefont {Andersen}}]{Pavarini2004}%
  \BibitemOpen
  \bibfield  {author} {\bibinfo {author} {\bibfnamefont {E.}~\bibnamefont {Pavarini}}, \bibinfo {author} {\bibfnamefont {S.}~\bibnamefont {Biermann}}, \bibinfo {author} {\bibfnamefont {A.}~\bibnamefont {Poteryaev}}, \bibinfo {author} {\bibfnamefont {A.~I.}\ \bibnamefont {Lichtenstein}}, \bibinfo {author} {\bibfnamefont {A.}~\bibnamefont {Georges}},\ and\ \bibinfo {author} {\bibfnamefont {O.~K.}\ \bibnamefont {Andersen}},\ }\bibfield  {title} {\bibinfo {title} {Mott transition and suppression of orbital fluctuations in orthorhombic $3{d}^{1}$ perovskites},\ }\href {https://doi.org/10.1103/PhysRevLett.92.176403} {\bibfield  {journal} {\bibinfo  {journal} {Phys. Rev. Lett.}\ }\textbf {\bibinfo {volume} {92}},\ \bibinfo {pages} {176403} (\bibinfo {year} {2004})}\BibitemShut {NoStop}%
\bibitem [{\citenamefont {Yoshida}\ \emph {et~al.}(2005)\citenamefont {Yoshida}, \citenamefont {Tanaka}, \citenamefont {Yagi}, \citenamefont {Ino}, \citenamefont {Eisaki}, \citenamefont {Fujimori},\ and\ \citenamefont {Shen}}]{srvo3_prl}%
  \BibitemOpen
  \bibfield  {author} {\bibinfo {author} {\bibfnamefont {T.}~\bibnamefont {Yoshida}}, \bibinfo {author} {\bibfnamefont {K.}~\bibnamefont {Tanaka}}, \bibinfo {author} {\bibfnamefont {H.}~\bibnamefont {Yagi}}, \bibinfo {author} {\bibfnamefont {A.}~\bibnamefont {Ino}}, \bibinfo {author} {\bibfnamefont {H.}~\bibnamefont {Eisaki}}, \bibinfo {author} {\bibfnamefont {A.}~\bibnamefont {Fujimori}},\ and\ \bibinfo {author} {\bibfnamefont {Z.-X.}\ \bibnamefont {Shen}},\ }\bibfield  {title} {\bibinfo {title} {Direct observation of the mass renormalization in ${\mathrm{srvo}}_{3}$ by angle resolved photoemission spectroscopy},\ }\href {https://doi.org/10.1103/PhysRevLett.95.146404} {\bibfield  {journal} {\bibinfo  {journal} {Phys. Rev. Lett.}\ }\textbf {\bibinfo {volume} {95}},\ \bibinfo {pages} {146404} (\bibinfo {year} {2005})}\BibitemShut {NoStop}%
\bibitem [{\citenamefont {Sakuma}\ \emph {et~al.}(2013)\citenamefont {Sakuma}, \citenamefont {Werner},\ and\ \citenamefont {Aryasetiawan}}]{Sakuma2013}%
  \BibitemOpen
  \bibfield  {author} {\bibinfo {author} {\bibfnamefont {R.}~\bibnamefont {Sakuma}}, \bibinfo {author} {\bibfnamefont {P.}~\bibnamefont {Werner}},\ and\ \bibinfo {author} {\bibfnamefont {F.}~\bibnamefont {Aryasetiawan}},\ }\bibfield  {title} {\bibinfo {title} {Electronic structure of srvo${}_{3}$ within $gw$+dmft},\ }\href {https://doi.org/10.1103/PhysRevB.88.235110} {\bibfield  {journal} {\bibinfo  {journal} {Phys. Rev. B}\ }\textbf {\bibinfo {volume} {88}},\ \bibinfo {pages} {235110} (\bibinfo {year} {2013})}\BibitemShut {NoStop}%
\bibitem [{\citenamefont {Tomczak}\ \emph {et~al.}(2014)\citenamefont {Tomczak}, \citenamefont {Casula}, \citenamefont {Miyake},\ and\ \citenamefont {Biermann}}]{TomczakSrvo3}%
  \BibitemOpen
  \bibfield  {author} {\bibinfo {author} {\bibfnamefont {J.~M.}\ \bibnamefont {Tomczak}}, \bibinfo {author} {\bibfnamefont {M.}~\bibnamefont {Casula}}, \bibinfo {author} {\bibfnamefont {T.}~\bibnamefont {Miyake}},\ and\ \bibinfo {author} {\bibfnamefont {S.}~\bibnamefont {Biermann}},\ }\bibfield  {title} {\bibinfo {title} {Asymmetry in band widening and quasiparticle lifetimes in ${\mathrm{srvo}}_{3}$: Competition between screened exchange and local correlations from combined $gw$ and dynamical mean-field theory $gw + \mathrm{DMFT}$},\ }\href {https://doi.org/10.1103/PhysRevB.90.165138} {\bibfield  {journal} {\bibinfo  {journal} {Phys. Rev. B}\ }\textbf {\bibinfo {volume} {90}},\ \bibinfo {pages} {165138} (\bibinfo {year} {2014})}\BibitemShut {NoStop}%
\bibitem [{\citenamefont {Backes}\ \emph {et~al.}(2016)\citenamefont {Backes}, \citenamefont {R\"odel}, \citenamefont {Fortuna}, \citenamefont {Frantzeskakis}, \citenamefont {Le~F\`evre}, \citenamefont {Bertran}, \citenamefont {Kobayashi}, \citenamefont {Yukawa}, \citenamefont {Mitsuhashi}, \citenamefont {Kitamura}, \citenamefont {Horiba}, \citenamefont {Kumigashira}, \citenamefont {Saint-Martin}, \citenamefont {Fouchet}, \citenamefont {Berini}, \citenamefont {Dumont}, \citenamefont {Kim}, \citenamefont {Lechermann}, \citenamefont {Jeschke}, \citenamefont {Rozenberg}, \citenamefont {Valent\'{\i}},\ and\ \citenamefont {Santander-Syro}}]{BackesSrvo3}%
  \BibitemOpen
  \bibfield  {author} {\bibinfo {author} {\bibfnamefont {S.}~\bibnamefont {Backes}}, \bibinfo {author} {\bibfnamefont {T.~C.}\ \bibnamefont {R\"odel}}, \bibinfo {author} {\bibfnamefont {F.}~\bibnamefont {Fortuna}}, \bibinfo {author} {\bibfnamefont {E.}~\bibnamefont {Frantzeskakis}}, \bibinfo {author} {\bibfnamefont {P.}~\bibnamefont {Le~F\`evre}}, \bibinfo {author} {\bibfnamefont {F.}~\bibnamefont {Bertran}}, \bibinfo {author} {\bibfnamefont {M.}~\bibnamefont {Kobayashi}}, \bibinfo {author} {\bibfnamefont {R.}~\bibnamefont {Yukawa}}, \bibinfo {author} {\bibfnamefont {T.}~\bibnamefont {Mitsuhashi}}, \bibinfo {author} {\bibfnamefont {M.}~\bibnamefont {Kitamura}}, \bibinfo {author} {\bibfnamefont {K.}~\bibnamefont {Horiba}}, \bibinfo {author} {\bibfnamefont {H.}~\bibnamefont {Kumigashira}}, \bibinfo {author} {\bibfnamefont {R.}~\bibnamefont {Saint-Martin}}, \bibinfo {author} {\bibfnamefont {A.}~\bibnamefont {Fouchet}}, \bibinfo {author} {\bibfnamefont {B.}~\bibnamefont {Berini}}, \bibinfo {author} {\bibfnamefont {Y.}~\bibnamefont {Dumont}}, \bibinfo {author} {\bibfnamefont {A.~J.}\ \bibnamefont {Kim}}, \bibinfo {author} {\bibfnamefont {F.}~\bibnamefont {Lechermann}}, \bibinfo {author} {\bibfnamefont {H.~O.}\ \bibnamefont {Jeschke}}, \bibinfo {author} {\bibfnamefont {M.~J.}\ \bibnamefont {Rozenberg}}, \bibinfo {author} {\bibfnamefont {R.}~\bibnamefont {Valent\'{\i}}},\ and\ \bibinfo {author} {\bibfnamefont {A.~F.}\ \bibnamefont {Santander-Syro}},\ }\bibfield  {title} {\bibinfo {title} {Hubbard band versus oxygen vacancy states in the correlated electron metal ${\mathrm{srvo}}_{3}$},\ }\href {https://doi.org/10.1103/PhysRevB.94.241110} {\bibfield  {journal} {\bibinfo  {journal} {Phys. Rev. B}\ }\textbf {\bibinfo {volume} {94}},\ \bibinfo {pages} {241110} (\bibinfo {year} {2016})}\BibitemShut {NoStop}%
\bibitem [{\citenamefont {Nakamura}\ \emph {et~al.}(2016)\citenamefont {Nakamura}, \citenamefont {Nohara}, \citenamefont {Yosimoto},\ and\ \citenamefont {Nomura}}]{Nakamura2016}%
  \BibitemOpen
  \bibfield  {author} {\bibinfo {author} {\bibfnamefont {K.}~\bibnamefont {Nakamura}}, \bibinfo {author} {\bibfnamefont {Y.}~\bibnamefont {Nohara}}, \bibinfo {author} {\bibfnamefont {Y.}~\bibnamefont {Yosimoto}},\ and\ \bibinfo {author} {\bibfnamefont {Y.}~\bibnamefont {Nomura}},\ }\bibfield  {title} {\bibinfo {title} {Ab initio $gw$ plus cumulant calculation for isolated band systems: Application to organic conductor ${(\mathrm{TMTSF})}_{2}{\mathrm{pf}}_{6}$ and transition-metal oxide ${\mathrm{srvo}}_{3}$},\ }\href {https://doi.org/10.1103/PhysRevB.93.085124} {\bibfield  {journal} {\bibinfo  {journal} {Phys. Rev. B}\ }\textbf {\bibinfo {volume} {93}},\ \bibinfo {pages} {085124} (\bibinfo {year} {2016})}\BibitemShut {NoStop}%
\bibitem [{\citenamefont {Petocchi}\ \emph {et~al.}(2020)\citenamefont {Petocchi}, \citenamefont {Nilsson}, \citenamefont {Aryasetiawan},\ and\ \citenamefont {Werner}}]{PetocchiScreening}%
  \BibitemOpen
  \bibfield  {author} {\bibinfo {author} {\bibfnamefont {F.}~\bibnamefont {Petocchi}}, \bibinfo {author} {\bibfnamefont {F.}~\bibnamefont {Nilsson}}, \bibinfo {author} {\bibfnamefont {F.}~\bibnamefont {Aryasetiawan}},\ and\ \bibinfo {author} {\bibfnamefont {P.}~\bibnamefont {Werner}},\ }\bibfield  {title} {\bibinfo {title} {Screening from ${e}_{g}$ states and antiferromagnetic correlations in ${d}^{(1,2,3)}$ perovskites: A $gw+\text{EDMFT}$ investigation},\ }\href {https://doi.org/10.1103/PhysRevResearch.2.013191} {\bibfield  {journal} {\bibinfo  {journal} {Phys. Rev. Res.}\ }\textbf {\bibinfo {volume} {2}},\ \bibinfo {pages} {013191} (\bibinfo {year} {2020})}\BibitemShut {NoStop}%
\bibitem [{\citenamefont {Inoue}\ \emph {et~al.}(1994)\citenamefont {Inoue}, \citenamefont {Hase}, \citenamefont {Aiura}, \citenamefont {Fujimori}, \citenamefont {Morikawa}, \citenamefont {Mizokawa}, \citenamefont {Haruyama}, \citenamefont {Maruyama},\ and\ \citenamefont {Nishihara}}]{Inoue1994}%
  \BibitemOpen
  \bibfield  {author} {\bibinfo {author} {\bibfnamefont {I.}~\bibnamefont {Inoue}}, \bibinfo {author} {\bibfnamefont {I.}~\bibnamefont {Hase}}, \bibinfo {author} {\bibfnamefont {Y.}~\bibnamefont {Aiura}}, \bibinfo {author} {\bibfnamefont {A.}~\bibnamefont {Fujimori}}, \bibinfo {author} {\bibfnamefont {K.}~\bibnamefont {Morikawa}}, \bibinfo {author} {\bibfnamefont {T.}~\bibnamefont {Mizokawa}}, \bibinfo {author} {\bibfnamefont {Y.}~\bibnamefont {Haruyama}}, \bibinfo {author} {\bibfnamefont {T.}~\bibnamefont {Maruyama}},\ and\ \bibinfo {author} {\bibfnamefont {Y.}~\bibnamefont {Nishihara}},\ }\bibfield  {title} {\bibinfo {title} {Systematic change of spectral function observed by controlling electron correlation in ca1-xsrxvo3 with fixed 3d1 configuration.},\ }\href {https://doi.org/https://doi.org/10.1016/0921-4534(94)91728-0} {\bibfield  {journal} {\bibinfo  {journal} {Physica C: Superconductivity}\ }\textbf {\bibinfo {volume} {235-240}},\ \bibinfo {pages} {1007} (\bibinfo {year} {1994})}\BibitemShut {NoStop}%
\bibitem [{\citenamefont {Sch\"uler}\ \emph {et~al.}(2013)\citenamefont {Sch\"uler}, \citenamefont {R\"osner}, \citenamefont {Wehling}, \citenamefont {Lichtenstein},\ and\ \citenamefont {Katsnelson}}]{Schueler2013}%
  \BibitemOpen
  \bibfield  {author} {\bibinfo {author} {\bibfnamefont {M.}~\bibnamefont {Sch\"uler}}, \bibinfo {author} {\bibfnamefont {M.}~\bibnamefont {R\"osner}}, \bibinfo {author} {\bibfnamefont {T.~O.}\ \bibnamefont {Wehling}}, \bibinfo {author} {\bibfnamefont {A.~I.}\ \bibnamefont {Lichtenstein}},\ and\ \bibinfo {author} {\bibfnamefont {M.~I.}\ \bibnamefont {Katsnelson}},\ }\bibfield  {title} {\bibinfo {title} {Optimal hubbard models for materials with nonlocal coulomb interactions: Graphene, silicene, and benzene},\ }\href {https://doi.org/10.1103/PhysRevLett.111.036601} {\bibfield  {journal} {\bibinfo  {journal} {Phys. Rev. Lett.}\ }\textbf {\bibinfo {volume} {111}},\ \bibinfo {pages} {036601} (\bibinfo {year} {2013})}\BibitemShut {NoStop}%
\bibitem [{\citenamefont {Su}\ \emph {et~al.}(2023)\citenamefont {Su}, \citenamefont {Ruotsalainen}, \citenamefont {Nicolaou}, \citenamefont {Gatti},\ and\ \citenamefont {Gloter}}]{srvo3_eels_wiley}%
  \BibitemOpen
  \bibfield  {author} {\bibinfo {author} {\bibfnamefont {C.-P.}\ \bibnamefont {Su}}, \bibinfo {author} {\bibfnamefont {K.}~\bibnamefont {Ruotsalainen}}, \bibinfo {author} {\bibfnamefont {A.}~\bibnamefont {Nicolaou}}, \bibinfo {author} {\bibfnamefont {M.}~\bibnamefont {Gatti}},\ and\ \bibinfo {author} {\bibfnamefont {A.}~\bibnamefont {Gloter}},\ }\bibfield  {title} {\bibinfo {title} {Plasmonic properties of srvo3 bulk and nanostructures},\ }\href {https://doi.org/https://doi.org/10.1002/adom.202202415} {\bibfield  {journal} {\bibinfo  {journal} {Advanced Optical Materials}\ }\textbf {\bibinfo {volume} {11}},\ \bibinfo {pages} {2202415} (\bibinfo {year} {2023})}\BibitemShut {NoStop}%
\bibitem [{\citenamefont {van Wezel}\ \emph {et~al.}(2011)\citenamefont {van Wezel}, \citenamefont {Schuster}, \citenamefont {K\"onig}, \citenamefont {Knupfer}, \citenamefont {van~den Brink}, \citenamefont {Berger},\ and\ \citenamefont {B\"uchner}}]{plasmons_prl}%
  \BibitemOpen
  \bibfield  {author} {\bibinfo {author} {\bibfnamefont {J.}~\bibnamefont {van Wezel}}, \bibinfo {author} {\bibfnamefont {R.}~\bibnamefont {Schuster}}, \bibinfo {author} {\bibfnamefont {A.}~\bibnamefont {K\"onig}}, \bibinfo {author} {\bibfnamefont {M.}~\bibnamefont {Knupfer}}, \bibinfo {author} {\bibfnamefont {J.}~\bibnamefont {van~den Brink}}, \bibinfo {author} {\bibfnamefont {H.}~\bibnamefont {Berger}},\ and\ \bibinfo {author} {\bibfnamefont {B.}~\bibnamefont {B\"uchner}},\ }\bibfield  {title} {\bibinfo {title} {Effect of charge order on the plasmon dispersion in transition-metal dichalcogenides},\ }\href {https://doi.org/10.1103/PhysRevLett.107.176404} {\bibfield  {journal} {\bibinfo  {journal} {Phys. Rev. Lett.}\ }\textbf {\bibinfo {volume} {107}},\ \bibinfo {pages} {176404} (\bibinfo {year} {2011})}\BibitemShut {NoStop}%
\bibitem [{\citenamefont {Husain}\ \emph {et~al.}(2023)\citenamefont {Husain}, \citenamefont {Huang}, \citenamefont {Mitrano}, \citenamefont {Rak}, \citenamefont {Rubeck}, \citenamefont {Guo}, \citenamefont {Yang}, \citenamefont {Sow}, \citenamefont {Maeno}, \citenamefont {Uchoa}, \citenamefont {Chiang}, \citenamefont {Batson}, \citenamefont {Phillips},\ and\ \citenamefont {Abbamonte}}]{pines_nature}%
  \BibitemOpen
  \bibfield  {author} {\bibinfo {author} {\bibfnamefont {A.~A.}\ \bibnamefont {Husain}}, \bibinfo {author} {\bibfnamefont {E.~W.}\ \bibnamefont {Huang}}, \bibinfo {author} {\bibfnamefont {M.}~\bibnamefont {Mitrano}}, \bibinfo {author} {\bibfnamefont {M.~S.}\ \bibnamefont {Rak}}, \bibinfo {author} {\bibfnamefont {S.~I.}\ \bibnamefont {Rubeck}}, \bibinfo {author} {\bibfnamefont {X.}~\bibnamefont {Guo}}, \bibinfo {author} {\bibfnamefont {H.}~\bibnamefont {Yang}}, \bibinfo {author} {\bibfnamefont {C.}~\bibnamefont {Sow}}, \bibinfo {author} {\bibfnamefont {Y.}~\bibnamefont {Maeno}}, \bibinfo {author} {\bibfnamefont {B.}~\bibnamefont {Uchoa}}, \bibinfo {author} {\bibfnamefont {T.~C.}\ \bibnamefont {Chiang}}, \bibinfo {author} {\bibfnamefont {P.~E.}\ \bibnamefont {Batson}}, \bibinfo {author} {\bibfnamefont {P.~W.}\ \bibnamefont {Phillips}},\ and\ \bibinfo {author} {\bibfnamefont {P.}~\bibnamefont {Abbamonte}},\ }\bibfield  {title} {\bibinfo {title} {Pines'demon observed as a 3d acoustic plasmon in sr2ruo4},\ }\href {https://doi.org/10.1038/s41586-023-06318-8} {\bibfield  {journal} {\bibinfo  {journal} {Nature}\ }\textbf {\bibinfo {volume} {621}},\ \bibinfo {pages} {66} (\bibinfo {year} {2023})}\BibitemShut {NoStop}%
\bibitem [{\citenamefont {Chamberland}(1967)}]{chamberland}%
  \BibitemOpen
  \bibfield  {author} {\bibinfo {author} {\bibfnamefont {B.}~\bibnamefont {Chamberland}},\ }\bibfield  {title} {\bibinfo {title} {Preparation and properties of srcro3},\ }\href {https://doi.org/https://doi.org/10.1016/0038-1098(67)90088-9} {\bibfield  {journal} {\bibinfo  {journal} {Solid State Communications}\ }\textbf {\bibinfo {volume} {5}},\ \bibinfo {pages} {663} (\bibinfo {year} {1967})}\BibitemShut {NoStop}%
\bibitem [{\citenamefont {Zhou}\ \emph {et~al.}(2006)\citenamefont {Zhou}, \citenamefont {Jin}, \citenamefont {Long}, \citenamefont {Yang},\ and\ \citenamefont {Goodenough}}]{srcro3_zhou}%
  \BibitemOpen
  \bibfield  {author} {\bibinfo {author} {\bibfnamefont {J.-S.}\ \bibnamefont {Zhou}}, \bibinfo {author} {\bibfnamefont {C.-Q.}\ \bibnamefont {Jin}}, \bibinfo {author} {\bibfnamefont {Y.-W.}\ \bibnamefont {Long}}, \bibinfo {author} {\bibfnamefont {L.-X.}\ \bibnamefont {Yang}},\ and\ \bibinfo {author} {\bibfnamefont {J.~B.}\ \bibnamefont {Goodenough}},\ }\bibfield  {title} {\bibinfo {title} {Anomalous electronic state in ${\mathrm{cacro}}_{3}$ and ${\mathrm{srcro}}_{3}$},\ }\href {https://doi.org/10.1103/PhysRevLett.96.046408} {\bibfield  {journal} {\bibinfo  {journal} {Phys. Rev. Lett.}\ }\textbf {\bibinfo {volume} {96}},\ \bibinfo {pages} {046408} (\bibinfo {year} {2006})}\BibitemShut {NoStop}%
\bibitem [{\citenamefont {Alario-Franco}(2008)}]{srcro3_eels}%
  \BibitemOpen
  \bibfield  {author} {\bibinfo {author} {\bibnamefont {Alario-Franco}},\ }\bibfield  {title} {\bibinfo {title} {Electron energy loss spectroscopy in acro3 (a = ca, sr and pb) perovskites},\ }\href {https://doi.org/10.1088/0953-8984/20/50/505207} {\bibfield  {journal} {\bibinfo  {journal} {Journal of Physics: Condensed Matter}\ }\textbf {\bibinfo {volume} {20}},\ \bibinfo {pages} {505207} (\bibinfo {year} {2008})}\BibitemShut {NoStop}%
\bibitem [{\citenamefont {Qian}\ \emph {et~al.}(2011)\citenamefont {Qian}, \citenamefont {Wang}, \citenamefont {Li}, \citenamefont {Jin},\ and\ \citenamefont {Fang}}]{srcro3_qian}%
  \BibitemOpen
  \bibfield  {author} {\bibinfo {author} {\bibfnamefont {Y.}~\bibnamefont {Qian}}, \bibinfo {author} {\bibfnamefont {G.}~\bibnamefont {Wang}}, \bibinfo {author} {\bibfnamefont {Z.}~\bibnamefont {Li}}, \bibinfo {author} {\bibfnamefont {C.~Q.}\ \bibnamefont {Jin}},\ and\ \bibinfo {author} {\bibfnamefont {Z.}~\bibnamefont {Fang}},\ }\bibfield  {title} {\bibinfo {title} {The electronic structure of a weakly correlated antiferromagnetic metal, srcro3: first-principles calculations},\ }\href {https://doi.org/10.1088/1367-2630/13/5/053002} {\bibfield  {journal} {\bibinfo  {journal} {New Journal of Physics}\ }\textbf {\bibinfo {volume} {13}},\ \bibinfo {pages} {053002} (\bibinfo {year} {2011})}\BibitemShut {NoStop}%
\bibitem [{\citenamefont {Zhang}\ \emph {et~al.}(2015)\citenamefont {Zhang}, \citenamefont {Du}, \citenamefont {Sushko}, \citenamefont {Bowden}, \citenamefont {Shutthanandan}, \citenamefont {Qiao}, \citenamefont {Cao}, \citenamefont {Gai}, \citenamefont {Sallis}, \citenamefont {Piper},\ and\ \citenamefont {Chambers}}]{zhang_srcro3}%
  \BibitemOpen
  \bibfield  {author} {\bibinfo {author} {\bibfnamefont {K.~H.~L.}\ \bibnamefont {Zhang}}, \bibinfo {author} {\bibfnamefont {Y.}~\bibnamefont {Du}}, \bibinfo {author} {\bibfnamefont {P.~V.}\ \bibnamefont {Sushko}}, \bibinfo {author} {\bibfnamefont {M.~E.}\ \bibnamefont {Bowden}}, \bibinfo {author} {\bibfnamefont {V.}~\bibnamefont {Shutthanandan}}, \bibinfo {author} {\bibfnamefont {L.}~\bibnamefont {Qiao}}, \bibinfo {author} {\bibfnamefont {G.~X.}\ \bibnamefont {Cao}}, \bibinfo {author} {\bibfnamefont {Z.}~\bibnamefont {Gai}}, \bibinfo {author} {\bibfnamefont {S.}~\bibnamefont {Sallis}}, \bibinfo {author} {\bibfnamefont {L.~F.~J.}\ \bibnamefont {Piper}},\ and\ \bibinfo {author} {\bibfnamefont {S.~A.}\ \bibnamefont {Chambers}},\ }\bibfield  {title} {\bibinfo {title} {Electronic and magnetic properties of epitaxial perovskite srcro3(001)},\ }\href {https://doi.org/10.1088/0953-8984/27/24/245605} {\bibfield  {journal} {\bibinfo  {journal} {Journal of Physics: Condensed Matter}\ }\textbf {\bibinfo {volume} {27}},\ \bibinfo {pages} {245605} (\bibinfo {year} {2015})}\BibitemShut {NoStop}%
\bibitem [{\citenamefont {Carta}\ \emph {et~al.}(2023)\citenamefont {Carta}, \citenamefont {Panda},\ and\ \citenamefont {Ederer}}]{carta2023emergence}%
  \BibitemOpen
  \bibfield  {author} {\bibinfo {author} {\bibfnamefont {A.}~\bibnamefont {Carta}}, \bibinfo {author} {\bibfnamefont {A.}~\bibnamefont {Panda}},\ and\ \bibinfo {author} {\bibfnamefont {C.}~\bibnamefont {Ederer}},\ }\href@noop {} {\bibinfo {title} {Emergence of a potential charge disproportionated insulating state in srcro$_{3}$}} (\bibinfo {year} {2023}),\ \Eprint {https://arxiv.org/abs/2312.12033} {arXiv:2312.12033 [cond-mat.str-el]} \BibitemShut {NoStop}%
\bibitem [{\citenamefont {Chmaissem}\ \emph {et~al.}(2001)\citenamefont {Chmaissem}, \citenamefont {Dabrowski}, \citenamefont {Kolesnik}, \citenamefont {Mais}, \citenamefont {Brown}, \citenamefont {Kruk}, \citenamefont {Prior}, \citenamefont {Pyles},\ and\ \citenamefont {Jorgensen}}]{chmaissem_srmno3}%
  \BibitemOpen
  \bibfield  {author} {\bibinfo {author} {\bibfnamefont {O.}~\bibnamefont {Chmaissem}}, \bibinfo {author} {\bibfnamefont {B.}~\bibnamefont {Dabrowski}}, \bibinfo {author} {\bibfnamefont {S.}~\bibnamefont {Kolesnik}}, \bibinfo {author} {\bibfnamefont {J.}~\bibnamefont {Mais}}, \bibinfo {author} {\bibfnamefont {D.~E.}\ \bibnamefont {Brown}}, \bibinfo {author} {\bibfnamefont {R.}~\bibnamefont {Kruk}}, \bibinfo {author} {\bibfnamefont {P.}~\bibnamefont {Prior}}, \bibinfo {author} {\bibfnamefont {B.}~\bibnamefont {Pyles}},\ and\ \bibinfo {author} {\bibfnamefont {J.~D.}\ \bibnamefont {Jorgensen}},\ }\bibfield  {title} {\bibinfo {title} {Relationship between structural parameters and the n\'eel temperature in ${\mathrm{sr}}_{1\ensuremath{-}x}{\mathrm{ca}}_{x}{\mathrm{mno}}_{3}$ $(0 <~x <~1)$ and ${\mathrm{sr}}_{1\ensuremath{-}y}{\mathrm{ba}}_{y}{\mathrm{mno}}_{3}$ $(y <~0.2)$},\ }\href {https://doi.org/10.1103/PhysRevB.64.134412} {\bibfield  {journal} {\bibinfo  {journal} {Phys. Rev. B}\ }\textbf {\bibinfo {volume} {64}},\ \bibinfo {pages} {134412} (\bibinfo {year} {2001})}\BibitemShut {NoStop}%
\bibitem [{\citenamefont {Takeda}\ and\ \citenamefont {Ohara}(1974)}]{takeda_srmno3}%
  \BibitemOpen
  \bibfield  {author} {\bibinfo {author} {\bibfnamefont {T.}~\bibnamefont {Takeda}}\ and\ \bibinfo {author} {\bibfnamefont {S.}~\bibnamefont {Ohara}},\ }\bibfield  {title} {\bibinfo {title} {Magnetic structure of the cubic perovskite type srmno<sub>3</sub>},\ }\href {https://doi.org/10.1143/JPSJ.37.275} {\bibfield  {journal} {\bibinfo  {journal} {Journal of the Physical Society of Japan}\ }\textbf {\bibinfo {volume} {37}},\ \bibinfo {pages} {275} (\bibinfo {year} {1974})}\BibitemShut {NoStop}%
\bibitem [{\citenamefont {Shinaoka}\ \emph {et~al.}(2015)\citenamefont {Shinaoka}, \citenamefont {Troyer},\ and\ \citenamefont {Werner}}]{Shinaoka2015}%
  \BibitemOpen
  \bibfield  {author} {\bibinfo {author} {\bibfnamefont {H.}~\bibnamefont {Shinaoka}}, \bibinfo {author} {\bibfnamefont {M.}~\bibnamefont {Troyer}},\ and\ \bibinfo {author} {\bibfnamefont {P.}~\bibnamefont {Werner}},\ }\bibfield  {title} {\bibinfo {title} {Accuracy of downfolding based on the constrained random-phase approximation},\ }\href {https://doi.org/10.1103/PhysRevB.91.245156} {\bibfield  {journal} {\bibinfo  {journal} {Phys. Rev. B}\ }\textbf {\bibinfo {volume} {91}},\ \bibinfo {pages} {245156} (\bibinfo {year} {2015})}\BibitemShut {NoStop}%
\bibitem [{fle()}]{fleur}%
  \BibitemOpen
  \href {http://www.flapw.de} {}\bibinfo {note} {FLEUR project: http://www.flapw.de}\BibitemShut {NoStop}%
\bibitem [{\citenamefont {Friedrich}\ \emph {et~al.}(2010)\citenamefont {Friedrich}, \citenamefont {Bl\"ugel},\ and\ \citenamefont {Schindlmayr}}]{spex}%
  \BibitemOpen
  \bibfield  {author} {\bibinfo {author} {\bibfnamefont {C.}~\bibnamefont {Friedrich}}, \bibinfo {author} {\bibfnamefont {S.}~\bibnamefont {Bl\"ugel}},\ and\ \bibinfo {author} {\bibfnamefont {A.}~\bibnamefont {Schindlmayr}},\ }\bibfield  {title} {\bibinfo {title} {Efficient implementation of the $gw$ approximation within the all-electron flapw method},\ }\href {https://doi.org/10.1103/PhysRevB.81.125102} {\bibfield  {journal} {\bibinfo  {journal} {Phys. Rev. B}\ }\textbf {\bibinfo {volume} {81}},\ \bibinfo {pages} {125102} (\bibinfo {year} {2010})}\BibitemShut {NoStop}%
\bibitem [{\citenamefont {Pizzi}\ \emph {et~al.}(2020)\citenamefont {Pizzi}, \citenamefont {Vitale}, \citenamefont {Arita}, \citenamefont {Blügel}, \citenamefont {Freimuth}, \citenamefont {G{\'{e}}ranton}, \citenamefont {Gibertini}, \citenamefont {Gresch}, \citenamefont {Johnson}, \citenamefont {Koretsune}, \citenamefont {Iba{\~{n}}ez-Azpiroz}, \citenamefont {Lee}, \citenamefont {Lihm}, \citenamefont {Marchand}, \citenamefont {Marrazzo}, \citenamefont {Mokrousov}, \citenamefont {Mustafa}, \citenamefont {Nohara}, \citenamefont {Nomura}, \citenamefont {Paulatto}, \citenamefont {Ponc{\'{e}}}, \citenamefont {Ponweiser}, \citenamefont {Qiao}, \citenamefont {Thöle}, \citenamefont {Tsirkin}, \citenamefont {Wierzbowska}, \citenamefont {Marzari}, \citenamefont {Vanderbilt}, \citenamefont {Souza}, \citenamefont {Mostofi},\ and\ \citenamefont {Yates}}]{wannier90}%
  \BibitemOpen
  \bibfield  {author} {\bibinfo {author} {\bibfnamefont {G.}~\bibnamefont {Pizzi}}, \bibinfo {author} {\bibfnamefont {V.}~\bibnamefont {Vitale}}, \bibinfo {author} {\bibfnamefont {R.}~\bibnamefont {Arita}}, \bibinfo {author} {\bibfnamefont {S.}~\bibnamefont {Blügel}}, \bibinfo {author} {\bibfnamefont {F.}~\bibnamefont {Freimuth}}, \bibinfo {author} {\bibfnamefont {G.}~\bibnamefont {G{\'{e}}ranton}}, \bibinfo {author} {\bibfnamefont {M.}~\bibnamefont {Gibertini}}, \bibinfo {author} {\bibfnamefont {D.}~\bibnamefont {Gresch}}, \bibinfo {author} {\bibfnamefont {C.}~\bibnamefont {Johnson}}, \bibinfo {author} {\bibfnamefont {T.}~\bibnamefont {Koretsune}}, \bibinfo {author} {\bibfnamefont {J.}~\bibnamefont {Iba{\~{n}}ez-Azpiroz}}, \bibinfo {author} {\bibfnamefont {H.}~\bibnamefont {Lee}}, \bibinfo {author} {\bibfnamefont {J.-M.}\ \bibnamefont {Lihm}}, \bibinfo {author} {\bibfnamefont {D.}~\bibnamefont {Marchand}}, \bibinfo {author} {\bibfnamefont {A.}~\bibnamefont {Marrazzo}}, \bibinfo {author} {\bibfnamefont {Y.}~\bibnamefont {Mokrousov}}, \bibinfo {author} {\bibfnamefont {J.~I.}\ \bibnamefont {Mustafa}}, \bibinfo {author} {\bibfnamefont {Y.}~\bibnamefont {Nohara}}, \bibinfo {author} {\bibfnamefont {Y.}~\bibnamefont {Nomura}}, \bibinfo {author} {\bibfnamefont {L.}~\bibnamefont {Paulatto}}, \bibinfo {author} {\bibfnamefont {S.}~\bibnamefont {Ponc{\'{e}}}}, \bibinfo {author} {\bibfnamefont {T.}~\bibnamefont {Ponweiser}}, \bibinfo {author} {\bibfnamefont {J.}~\bibnamefont {Qiao}}, \bibinfo {author} {\bibfnamefont {F.}~\bibnamefont {Thöle}}, \bibinfo {author} {\bibfnamefont {S.~S.}\ \bibnamefont {Tsirkin}}, \bibinfo {author} {\bibfnamefont {M.}~\bibnamefont {Wierzbowska}}, \bibinfo {author} {\bibfnamefont {N.}~\bibnamefont {Marzari}}, \bibinfo {author} {\bibfnamefont {D.}~\bibnamefont {Vanderbilt}}, \bibinfo {author} {\bibfnamefont {I.}~\bibnamefont {Souza}}, \bibinfo {author} {\bibfnamefont {A.~A.}\ \bibnamefont {Mostofi}},\ and\ \bibinfo {author} {\bibfnamefont {J.~R.}\ \bibnamefont {Yates}},\ }\bibfield  {title} {\bibinfo {title} {Wannier90 as a community code: new features and applications},\ }\href {https://doi.org/10.1088/1361-648x/ab51ff} {\bibfield  {journal} {\bibinfo  {journal} {Journal of Physics: Condensed Matter}\ }\textbf {\bibinfo {volume} {32}},\ \bibinfo {pages} {165902} (\bibinfo {year} {2020})}\BibitemShut {NoStop}%
\bibitem [{\citenamefont {Werner}\ \emph {et~al.}(2006)\citenamefont {Werner}, \citenamefont {Comanac}, \citenamefont {de' Medici}, \citenamefont {Troyer},\ and\ \citenamefont {Millis}}]{ct_hyb}%
  \BibitemOpen
  \bibfield  {author} {\bibinfo {author} {\bibfnamefont {P.}~\bibnamefont {Werner}}, \bibinfo {author} {\bibfnamefont {A.}~\bibnamefont {Comanac}}, \bibinfo {author} {\bibfnamefont {L.}~\bibnamefont {de' Medici}}, \bibinfo {author} {\bibfnamefont {M.}~\bibnamefont {Troyer}},\ and\ \bibinfo {author} {\bibfnamefont {A.~J.}\ \bibnamefont {Millis}},\ }\bibfield  {title} {\bibinfo {title} {Continuous-time solver for quantum impurity models},\ }\href {https://doi.org/10.1103/PhysRevLett.97.076405} {\bibfield  {journal} {\bibinfo  {journal} {Phys. Rev. Lett.}\ }\textbf {\bibinfo {volume} {97}},\ \bibinfo {pages} {076405} (\bibinfo {year} {2006})}\BibitemShut {NoStop}%
\bibitem [{\citenamefont {Werner}\ and\ \citenamefont {Millis}(2010)}]{ct_hyb_retarded}%
  \BibitemOpen
  \bibfield  {author} {\bibinfo {author} {\bibfnamefont {P.}~\bibnamefont {Werner}}\ and\ \bibinfo {author} {\bibfnamefont {A.~J.}\ \bibnamefont {Millis}},\ }\bibfield  {title} {\bibinfo {title} {Dynamical screening in correlated electron materials},\ }\href {https://doi.org/10.1103/PhysRevLett.104.146401} {\bibfield  {journal} {\bibinfo  {journal} {Phys. Rev. Lett.}\ }\textbf {\bibinfo {volume} {104}},\ \bibinfo {pages} {146401} (\bibinfo {year} {2010})}\BibitemShut {NoStop}%
\end{thebibliography}%


\begin{thebibliography}{13}%
\makeatletter
\providecommand \@ifxundefined [1]{%
 \@ifx{#1\undefined}
}%
\providecommand \@ifnum [1]{%
 \ifnum #1\expandafter \@firstoftwo
 \else \expandafter \@secondoftwo
 \fi
}%
\providecommand \@ifx [1]{%
 \ifx #1\expandafter \@firstoftwo
 \else \expandafter \@secondoftwo
 \fi
}%
\providecommand \natexlab [1]{#1}%
\providecommand \enquote  [1]{``#1''}%
\providecommand \bibnamefont  [1]{#1}%
\providecommand \bibfnamefont [1]{#1}%
\providecommand \citenamefont [1]{#1}%
\providecommand \href@noop [0]{\@secondoftwo}%
\providecommand \href [0]{\begingroup \@sanitize@url \@href}%
\providecommand \@href[1]{\@@startlink{#1}\@@href}%
\providecommand \@@href[1]{\endgroup#1\@@endlink}%
\providecommand \@sanitize@url [0]{\catcode `\\12\catcode `\$12\catcode `\&12\catcode `\#12\catcode `\^12\catcode `\_12\catcode `\%12\relax}%
\providecommand \@@startlink[1]{}%
\providecommand \@@endlink[0]{}%
\providecommand \url  [0]{\begingroup\@sanitize@url \@url }%
\providecommand \@url [1]{\endgroup\@href {#1}{\urlprefix }}%
\providecommand \urlprefix  [0]{URL }%
\providecommand \Eprint [0]{\href }%
\providecommand \doibase [0]{https://doi.org/}%
\providecommand \selectlanguage [0]{\@gobble}%
\providecommand \bibinfo  [0]{\@secondoftwo}%
\providecommand \bibfield  [0]{\@secondoftwo}%
\providecommand \translation [1]{[#1]}%
\providecommand \BibitemOpen [0]{}%
\providecommand \bibitemStop [0]{}%
\providecommand \bibitemNoStop [0]{.\EOS\space}%
\providecommand \EOS [0]{\spacefactor3000\relax}%
\providecommand \BibitemShut  [1]{\csname bibitem#1\endcsname}%
\let\auto@bib@innerbib\@empty
\bibitem [{\citenamefont {Hedin}(1965)}]{HedinGW}%
  \BibitemOpen
  \bibfield  {author} {\bibinfo {author} {\bibfnamefont {L.}~\bibnamefont {Hedin}},\ }\bibfield  {title} {\bibinfo {title} {New method for calculating the one-particle green's function with application to the electron-gas problem},\ }\href {https://doi.org/10.1103/PhysRev.139.A796} {\bibfield  {journal} {\bibinfo  {journal} {Phys. Rev.}\ }\textbf {\bibinfo {volume} {139}},\ \bibinfo {pages} {A796} (\bibinfo {year} {1965})}\BibitemShut {NoStop}%
\bibitem [{\citenamefont {Nilsson}\ \emph {et~al.}(2017)\citenamefont {Nilsson}, \citenamefont {Boehnke}, \citenamefont {Werner},\ and\ \citenamefont {Aryasetiawan}}]{NilssonGW_EDMFT}%
  \BibitemOpen
  \bibfield  {author} {\bibinfo {author} {\bibfnamefont {F.}~\bibnamefont {Nilsson}}, \bibinfo {author} {\bibfnamefont {L.}~\bibnamefont {Boehnke}}, \bibinfo {author} {\bibfnamefont {P.}~\bibnamefont {Werner}},\ and\ \bibinfo {author} {\bibfnamefont {F.}~\bibnamefont {Aryasetiawan}},\ }\bibfield  {title} {\bibinfo {title} {Multitier self-consistent $gw+\text{EDMFT}$},\ }\href {https://doi.org/10.1103/PhysRevMaterials.1.043803} {\bibfield  {journal} {\bibinfo  {journal} {Phys. Rev. Mater.}\ }\textbf {\bibinfo {volume} {1}},\ \bibinfo {pages} {043803} (\bibinfo {year} {2017})}\BibitemShut {NoStop}%
\bibitem [{\citenamefont {Werner}\ and\ \citenamefont {Casula}(2016)}]{WernerReview}%
  \BibitemOpen
  \bibfield  {author} {\bibinfo {author} {\bibfnamefont {P.}~\bibnamefont {Werner}}\ and\ \bibinfo {author} {\bibfnamefont {M.}~\bibnamefont {Casula}},\ }\bibfield  {title} {\bibinfo {title} {Dynamical screening in correlated electron systems---from lattice models to realistic materials},\ }\href {https://doi.org/10.1088/0953-8984/28/38/383001} {\bibfield  {journal} {\bibinfo  {journal} {Journal of Physics: Condensed Matter}\ }\textbf {\bibinfo {volume} {28}},\ \bibinfo {pages} {383001} (\bibinfo {year} {2016})}\BibitemShut {NoStop}%
\bibitem [{\citenamefont {Friedrich}\ \emph {et~al.}(2010)\citenamefont {Friedrich}, \citenamefont {Bl\"ugel},\ and\ \citenamefont {Schindlmayr}}]{spex}%
  \BibitemOpen
  \bibfield  {author} {\bibinfo {author} {\bibfnamefont {C.}~\bibnamefont {Friedrich}}, \bibinfo {author} {\bibfnamefont {S.}~\bibnamefont {Bl\"ugel}},\ and\ \bibinfo {author} {\bibfnamefont {A.}~\bibnamefont {Schindlmayr}},\ }\bibfield  {title} {\bibinfo {title} {Efficient implementation of the $gw$ approximation within the all-electron flapw method},\ }\href {https://doi.org/10.1103/PhysRevB.81.125102} {\bibfield  {journal} {\bibinfo  {journal} {Phys. Rev. B}\ }\textbf {\bibinfo {volume} {81}},\ \bibinfo {pages} {125102} (\bibinfo {year} {2010})}\BibitemShut {NoStop}%
\bibitem [{\citenamefont {Aryasetiawan}\ \emph {et~al.}(2004)\citenamefont {Aryasetiawan}, \citenamefont {Imada}, \citenamefont {Georges}, \citenamefont {Kotliar}, \citenamefont {Biermann},\ and\ \citenamefont {Lichtenstein}}]{Aryasetiawan2004}%
  \BibitemOpen
  \bibfield  {author} {\bibinfo {author} {\bibfnamefont {F.}~\bibnamefont {Aryasetiawan}}, \bibinfo {author} {\bibfnamefont {M.}~\bibnamefont {Imada}}, \bibinfo {author} {\bibfnamefont {A.}~\bibnamefont {Georges}}, \bibinfo {author} {\bibfnamefont {G.}~\bibnamefont {Kotliar}}, \bibinfo {author} {\bibfnamefont {S.}~\bibnamefont {Biermann}},\ and\ \bibinfo {author} {\bibfnamefont {A.~I.}\ \bibnamefont {Lichtenstein}},\ }\bibfield  {title} {\bibinfo {title} {Frequency-dependent local interactions and low-energy effective models from electronic structure calculations},\ }\href {https://doi.org/10.1103/PhysRevB.70.195104} {\bibfield  {journal} {\bibinfo  {journal} {Phys. Rev. B}\ }\textbf {\bibinfo {volume} {70}},\ \bibinfo {pages} {195104} (\bibinfo {year} {2004})}\BibitemShut {NoStop}%
\bibitem [{\citenamefont {Giuliani}\ and\ \citenamefont {Vignale}(2005)}]{vignale}%
  \BibitemOpen
  \bibfield  {author} {\bibinfo {author} {\bibfnamefont {G.}~\bibnamefont {Giuliani}}\ and\ \bibinfo {author} {\bibfnamefont {G.}~\bibnamefont {Vignale}},\ }\href {https://doi.org/10.1017/CBO9780511619915} {\emph {\bibinfo {title} {Quantum Theory of the Electron Liquid}}}\ (\bibinfo  {publisher} {Cambridge University Press},\ \bibinfo {year} {2005})\BibitemShut {NoStop}%
\bibitem [{\citenamefont {Gubernatis}\ \emph {et~al.}(1991)\citenamefont {Gubernatis}, \citenamefont {Jarrell}, \citenamefont {Silver},\ and\ \citenamefont {Sivia}}]{Gubernatis1991}%
  \BibitemOpen
  \bibfield  {author} {\bibinfo {author} {\bibfnamefont {J.~E.}\ \bibnamefont {Gubernatis}}, \bibinfo {author} {\bibfnamefont {M.}~\bibnamefont {Jarrell}}, \bibinfo {author} {\bibfnamefont {R.~N.}\ \bibnamefont {Silver}},\ and\ \bibinfo {author} {\bibfnamefont {D.~S.}\ \bibnamefont {Sivia}},\ }\bibfield  {title} {\bibinfo {title} {Quantum monte carlo simulations and maximum entropy: Dynamics from imaginary-time data},\ }\href {https://doi.org/10.1103/PhysRevB.44.6011} {\bibfield  {journal} {\bibinfo  {journal} {Phys. Rev. B}\ }\textbf {\bibinfo {volume} {44}},\ \bibinfo {pages} {6011} (\bibinfo {year} {1991})}\BibitemShut {NoStop}%
\bibitem [{\citenamefont {Egerton}(2011)}]{eels_bible}%
  \BibitemOpen
  \bibfield  {author} {\bibinfo {author} {\bibfnamefont {R.}~\bibnamefont {Egerton}},\ }\href {https://doi.org/https://doi.org/10.1007/978-1-4419-9583-4} {\emph {\bibinfo {title} {Electron Energy-Loss Spectroscopy in the Electron Microscope}}}\ (\bibinfo  {publisher} {Springer New York, NY},\ \bibinfo {year} {2011})\BibitemShut {NoStop}%
\bibitem [{\citenamefont {Del~Sole}\ and\ \citenamefont {Girlanda}(1993)}]{sole_eels}%
  \BibitemOpen
  \bibfield  {author} {\bibinfo {author} {\bibfnamefont {R.}~\bibnamefont {Del~Sole}}\ and\ \bibinfo {author} {\bibfnamefont {R.}~\bibnamefont {Girlanda}},\ }\bibfield  {title} {\bibinfo {title} {Optical properties of semiconductors within the independent-quasiparticle approximation},\ }\href {https://doi.org/10.1103/PhysRevB.48.11789} {\bibfield  {journal} {\bibinfo  {journal} {Phys. Rev. B}\ }\textbf {\bibinfo {volume} {48}},\ \bibinfo {pages} {11789} (\bibinfo {year} {1993})}\BibitemShut {NoStop}%
\bibitem [{\citenamefont {Kubo}(2007)}]{flex}%
  \BibitemOpen
  \bibfield  {author} {\bibinfo {author} {\bibfnamefont {K.}~\bibnamefont {Kubo}},\ }\bibfield  {title} {\bibinfo {title} {Pairing symmetry in a two-orbital hubbard model on a square lattice},\ }\href {https://doi.org/10.1103/PhysRevB.75.224509} {\bibfield  {journal} {\bibinfo  {journal} {Phys. Rev. B}\ }\textbf {\bibinfo {volume} {75}},\ \bibinfo {pages} {224509} (\bibinfo {year} {2007})}\BibitemShut {NoStop}%
\bibitem [{\citenamefont {Witt}\ \emph {et~al.}(2023)\citenamefont {Witt}, \citenamefont {Si}, \citenamefont {Tomczak}, \citenamefont {Held},\ and\ \citenamefont {Wehling}}]{no_sc}%
  \BibitemOpen
  \bibfield  {author} {\bibinfo {author} {\bibfnamefont {N.}~\bibnamefont {Witt}}, \bibinfo {author} {\bibfnamefont {L.}~\bibnamefont {Si}}, \bibinfo {author} {\bibfnamefont {J.~M.}\ \bibnamefont {Tomczak}}, \bibinfo {author} {\bibfnamefont {K.}~\bibnamefont {Held}},\ and\ \bibinfo {author} {\bibfnamefont {T.~O.}\ \bibnamefont {Wehling}},\ }\bibfield  {title} {\bibinfo {title} {{No superconductivity in Pb$_9$Cu$_1$(PO$_4$)$_6$O found in orbital and spin fluctuation exchange calculations}},\ }\href {https://doi.org/10.21468/SciPostPhys.15.5.197} {\bibfield  {journal} {\bibinfo  {journal} {SciPost Phys.}\ }\textbf {\bibinfo {volume} {15}},\ \bibinfo {pages} {197} (\bibinfo {year} {2023})}\BibitemShut {NoStop}%
\bibitem [{\citenamefont {{Y.M. Vilk}}\ and\ \citenamefont {{A.-M.S. Tremblay}}(1997)}]{vilk1997}%
  \BibitemOpen
  \bibfield  {author} {\bibinfo {author} {\bibnamefont {{Y.M. Vilk}}}\ and\ \bibinfo {author} {\bibnamefont {{A.-M.S. Tremblay}}},\ }\bibfield  {title} {\bibinfo {title} {Non-perturbative many-body approach to the hubbard model and single-particle pseudogap},\ }\href {https://doi.org/10.1051/jp1:1997135} {\bibfield  {journal} {\bibinfo  {journal} {J. Phys. I France}\ }\textbf {\bibinfo {volume} {7}},\ \bibinfo {pages} {1309} (\bibinfo {year} {1997})}\BibitemShut {NoStop}%
\bibitem [{\citenamefont {Simard}\ and\ \citenamefont {Werner}(2023)}]{simard2023}%
  \BibitemOpen
  \bibfield  {author} {\bibinfo {author} {\bibfnamefont {O.}~\bibnamefont {Simard}}\ and\ \bibinfo {author} {\bibfnamefont {P.}~\bibnamefont {Werner}},\ }\bibfield  {title} {\bibinfo {title} {Dynamical mean field theory extension to the nonequilibrium two-particle self-consistent approach},\ }\href {https://doi.org/10.1103/PhysRevB.107.245137} {\bibfield  {journal} {\bibinfo  {journal} {Phys. Rev. B}\ }\textbf {\bibinfo {volume} {107}},\ \bibinfo {pages} {245137} (\bibinfo {year} {2023})}\BibitemShut {NoStop}%
\end{thebibliography}%

\end{document}


\preprint{APS/123-QED}

\title{Internal consistency of multi-tier $GW$+EDMFT -- Supplementary Material}

\author{Ruslan Mushkaev}
\affiliation{Department of Physics, University of Fribourg, 1700 Fribourg, Switzerland}
\author{Francesco Petocchi}%
\affiliation{Department of Quantum Matter Physics, University of Geneva, 1211 Geneva 4, Switzerland}
\author{Viktor Christiansson}%
\affiliation{Department of Physics, University of Fribourg, 1700 Fribourg, Switzerland}
\author{Philipp Werner}%
\affiliation{Department of Physics, University of Fribourg, 1700 Fribourg, Switzerland}

\date{\today}

\maketitle
\onecolumngrid

\section{\label{gw_edmft} $GW$+EDMFT formalism}

\subsection{\label{GW} $GW$ approximation}

The self-energy $\Sigma$ encodes how correlations affect the propagation of electrons in a solid. To avoid problems with diverging self-energy diagrams when expanding in terms of the bare Coulomb interaction $v$, Hedin \cite{HedinGW} introduced an expansion in powers of the \textit{screened} interaction $W$. The first two terms of the self-energy expansion read:
\begin{align}
\label{sigma}
    \Sigma(1,2) = i G(1,2)W(1^+,2)
    - \int G(1,3) G(3,4) G(4,2) W(1,4) W(3,2) d3 d4 + \ldots,
\end{align}
where $G$ is the interacting electron Green's function and $(i,j)$ are space-time indices. Keeping only the leading order term $\Sigma(1,2) = i G(1,2)W(1^+,2)$ corresponds to the well-known $GW$ approximation, which partially captures non-local correlations via polarization effects. This can be seen by noting that $W$ obeys a bosonic Dyson equation
\begin{eqnarray}
\label{W}
    W(1,2) = v(1,2) + \int v(1,3) \Pi(3,4) W(4,2) d3 d4, 
\end{eqnarray}
where $\Pi(1,2)$ is the electron polarization function which in the Random Phase Approximation can be obtained as 
\begin{eqnarray}
\label{P}
    \Pi(1,2) = -i G(1,2)G(2,1^+).
\end{eqnarray}
Equations \eqref{W} and \eqref{P}, plus the first order term in Eq.~\eqref{sigma} and the corresponding Fermionic Dyson equation $G(1,2) = G_0(1,2) + \int G_0(1,3) \Sigma(3,4) G(4,2) d3 d4$ (with $G_0$ the noninteracting Green's function) define the $GW$ self-consistency loop. In practical first principles calculations, one replaces the exchange correlation potential in the DFT Kohn-Sham propagators by the single-shot or self-consistent (quasi-particle) $GW$ self-energy, which has been shown to yield improved descriptions of weakly and moderately correlated materials.

\subsection{\label{EDMFT} Extended Dynamical Mean Field Theory}

Within (E)DMFT, correlations are assumed to be local (i.e. the self-energy and polarization are momentum independent) and the system is self-consistently mapped onto an impurity model with a bath Green's function $\mathcal{G}$ (or hybridization function $\Delta$) and an effective dynamically screened bare interaction $\mathcal{U}$ \cite{NilssonGW_EDMFT,WernerReview}. $\mathcal{G}$ and $\mathcal{U}$ are determined self-consistently, by imposing that the physical impurity correlation functions are equal to their local lattice counterparts: $G^{\mathrm{imp}} = \frac{1}{N_k}\sum_k G^{\mathrm{lat}}(k) $ and $ W^{\mathrm{imp}} = \frac{1}{N_k}\sum_k W^{\mathrm{lat}}(k)$. The EDMFT self-consistency loop contains the following steps:

\begin{enumerate}
    \item Make an initial guess for the impurity self-energy and polarization $\Sigma^{\mathrm{imp}}$, $\Pi^{\mathrm{imp}}$.
    \item Identify the impurity self-energy and polarization with the corresponding lattice quantities (correlations are assumed to be local within EDMFT): $\Sigma^{\mathrm{lat}}=\Sigma^{\mathrm{imp}},\Pi^{\mathrm{lat}}=\Pi^{\mathrm{imp}}$.
    \item Compute the local lattice Green's function and screened interaction: 
    \begin{itemize}[leftmargin=-.15in]
        \item[] $G^{\mathrm{loc}} = \frac{1}{N_k}\sum_k G^{\mathrm{lat}}(k) = \frac{1}{N_k}\sum_k (G_0^{-1}(k) - \Sigma^{\mathrm{lat}} )^{-1}$,
        \item[] $W^{\mathrm{loc}} = \frac{1}{N_q}\sum_q W^{\mathrm{lat}}(q) = \frac{1}{N_q}\sum_q v(q) (1-\Pi^{\mathrm{lat}} v(q))^{-1}$.
    \end{itemize} 
    \item Impose the self consistency conditions: $G^{\mathrm{imp}} = G^{\mathrm{loc}}$ and $W^{\mathrm{imp}} = W^{\mathrm{loc}}$.
    \item Compute the fermionic and bosonic dynamical mean fields: 
    \begin{itemize}
    \item[] $\mathcal{G} = ({\gimp}^{-1} + \simp)^{-1}$, \item[] $\mathcal{U} = \wimp (1 + \pimp \wimp )^{-1}$.
    \end{itemize} 
    \item Solve the impurity problem with these dynamical mean fields, i.e., calculate the impurity Green's function $\gimp$ and the charge susceptibility $\cimp$.
    \item Update the impurity self-energy and polarization: 
        \begin{itemize}
    \item[] $\simp = \mathcal{G}^{-1} - {\gimp}^{-1}$,
    \item[] $\pimp = \cimp (\mathcal{U} \cimp -1)^{-1}$.
        \end{itemize} 
    \item Go back to step 2 until convergence is reached. 
\end{enumerate}

\subsection{\label{GW_EDMFT} Multi-tier $GW$+EDMFT}

$GW$+EDMFT considers a self-energy and polarization with both local and nonlocal components. Due to the separability of the vertex function in the Hedin equations \cite{NilssonGW_EDMFT}, the self-energy and polarization can be written as sums of their $GW$ parts and the vertex corrections:
$ \Sigma = \Sigma^{GW} + \Sigma^\textrm{vc}$ and $\Pi = \Pi^{GW} + \Pi^\textrm{vc}$. 
The vertex corrections are approximated by EDMFT: $\Sigma^\textrm{vc} \rightarrow \Sigma^{\mathrm{EMDFT}}$ and $\Pi^\textrm{vc} \rightarrow \Pi^{\mathrm{EMDFT}}$. Since $GW$ already contains local self-energy and polarization components, which are more accurately described by EDMFT, we subtract these components to avoid a double counting. This yields the following corrections to the momentum dependent self-energy and polarization:

\begin{align}
    &\Sigma^{GW+\text{EDMFT}}(k) = \Sigma^{GW}(k) - \frac{1}{N_k}\sum_k \Sigma^{GW}(k) 
    + \Sigma^\text{EDMFT},  \label{dc1} \\ 
    & \Pi^{GW+\text{EDMFT}}(q) = \Pi^{GG}(q) -\frac{1}{N_q}\sum_q \Pi^{GG}(q)  
    + \Pi^\text{EDMFT},
    \label{dc2}
\end{align}
where $\Sigma^{GW}$ and $\Pi^{GG}$ are defined in Sec.~\ref{GW}. The self-consistency cycle is equivalent to the one described in Sec.~\ref{EDMFT}, with the replacement of $\slat,\plat$ by Eqs.~\eqref{dc1} and \eqref{dc2}.

\begin{figure}
    \begin{centering}
\includegraphics[width=0.5\textwidth]{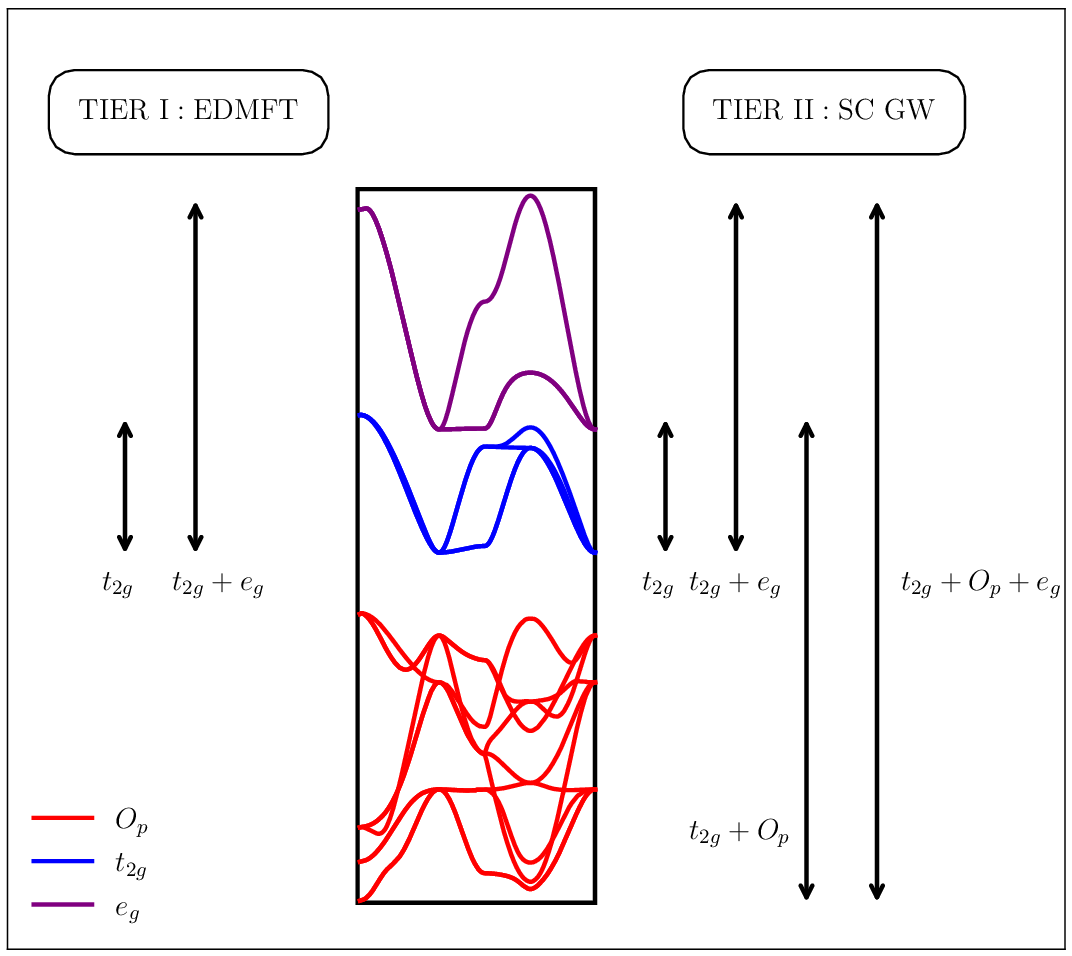}
\end{centering}
    \caption{Different model spaces used in the $GW$+EDMFT simulations of the cubic Perovskites. The most correlated $t_{2g}$ and $t_{2g}+e_g$ local subspaces are treated with EDMFT, while their non-local parts, as well as the lower lying Oxygen $p$-states are treated by self-consistent $GW$.}
    \label{model_spaces}
\end{figure}

In the multi-tier scheme, different energy windows are treated at different levels of theory. In the ``full" space (Tier III),  composed of $\sim 100$ bands, we consider the one-shot self-energy $\Sigma^{G^0W^0}$ and polarization $\Pi^{G^0G^0}$. The intermediate space (Tier II), spanning typically $\sim 10$ bands, is solved with self-consistent $GW$. Within the correlated space (Tier I), which can be equal to or smaller than the intermediate space, we treat the local correlations with EDMFT. The Tier I and Tier II spaces considered in the current study are illustrated in  Fig.~\ref{model_spaces}. The Tier III subspace corresponds to the $d$-orbital space. In practice, the basis used to describe the intermediate and correlated spaces is that of Maximally Localized Wannier Functions (MLWFs), obtained by Wannierizing an appropriate low-energy sector of the DFT bands. The one-shot $\Sigma^{G^0W^0}$ and $\Pi^{G^0G^0}$ are also rotated to this basis and define the ``bare" propagators and interactions in Tier II. 

\subsection{\label{g0w0_crpa} One-shot $G^0W^0$ and Constrained Random Phase Approximation}

We perform a one-shot $G^0W^0$ calculation in the TIER III space encompassing $\sim 100$ bands with the SPEX code \cite{spex} and use the resulting Green's functions as the bare propagators of the intermediate TIER II space \cite{NilssonGW_EDMFT}. 

The effective frequency- and momentum-dependent bare interaction of the intermediate space is computed within the constrained Random Phase Approximation (cRPA) \cite{Aryasetiawan2004}, which excludes screening processes occurring within the target space.

\section{Susceptibility calculations}

\subsection{Density-density response function \label{chi}}

The density-density correlation function $\chi_{n_q n_{-q}}(t)$, defined by \cite{vignale}
\begin{eqnarray}
    \chi_{n_q,n_{-q}}(t) = -i \lim_{\eta \rightarrow 0^+} \int_0^\infty \langle [n_q(t),n_{-q}]\rangle e^{i(\omega+i \eta)t} dt \,
    \label{charge_susc}
\end{eqnarray}
probes the charge excitations of a system and in the noninteracting case is related to the polarization function
\begin{equation}
    \Pi_{ijkl}(q,i \nu_m) = \frac{1}{\beta} \sum_{i\omega_n, k} G_{ik}(k,i\omega_n)G_{lj}(k-q,i\omega_n -i \nu_m)
    \label{pol}
\end{equation}
via
\begin{eqnarray}
    \chi_{n_q,n_{-q}}(q,\omega) = \sum_{ij} \Pi_{iijj}(q,\omega),
\end{eqnarray}
where $i,j$ are orbital indices.

The polarization function probes the density response to the total potential, $\Pi(1,2) \sim \frac{\delta n(1)}{\delta V_{\mathrm{tot}}(2)}$, where $V_{\mathrm{tot}}=V_{\mathrm{ext}}+V_{\mathrm{ind}}$ (with $V_{\mathrm{ind}}$ the Hartree potential), while the physically measurable charge susceptibility is the density response to the external potential, $\chi(1,2) \sim \frac{\delta n(1)}{\delta V_{\mathrm{ext}}(2)}$.
The two are related via $\chi = \Pi +\Pi v \chi$, where $v$ is the bare Coulomb interaction.

Within the $GW$+EDMFT formalism the polarization function $\Pi$ is defined in Eq.~\eqref{dc2} and the effective bare Coulomb interaction is taken to be the cRPA $U$, yielding 
\begin{align}
    \chi^{GW + \mathrm{EDMFT}}(q,i \nu_n) = \Pi(q,i \nu_n) [1-U^{\mathrm{cRPA}}(q,i \nu_n)\Pi(q,i \nu_n)]^{-1}.
    \label{chi_gw_dmft}
\end{align}
Here, the contraction of rank-4 tensor quantities is implemented as a matrix multiplication in a product basis of Wannier functions \cite{NilssonGW_EDMFT}. The full scalar charge susceptibility is then computed as $\chi^{GW + \mathrm{EDMFT}}_c(q,i \nu_n) = \sum_{i,j} \chi^{GW + \mathrm{EDMFT}}_{ii,jj}(q,i \nu_n)$. This function can be analytically continued with the Maximum Entropy (MaxEnt) method \cite{Gubernatis1991}.

\subsection{Dielectric function}

The matrix denominator in the expression for $\chi$ corresponds to the multi-orbital dielectric function $\epsilon=1-v \Pi$. This function can be probed by Electron Energy Loss Spectroscopy (EELS), which measures a related quantity $-\mathrm{Im} \frac{1}{\epsilon}$, known as the loss function \cite{eels_bible}. The experimentally measured dielectric function typically probes all inter- and intra-band transitions. Evaluated with Fermi's golden rule, its imaginary part reads \cite{sole_eels}
\begin{align}
    & \mathrm{Im} [\epsilon(q,\omega)] \sim \frac{4 \pi^2 e^2}{q^2}  \sum_{\alpha,\beta,k} |\langle \alpha,k+q|e^{i q \hat{r}} |\beta,k \rangle |^2 (1-f(E_{\alpha,k+q}))f(E_{\beta,k}) \delta(E_{\alpha,k+q}-E_{\beta,k}-\omega)
\end{align}
where $\alpha,\beta$ index the energy bands, $f(E)$ is the Fermi-Dirac distribution function and $v$ is the scalar bare Coulomb interaction $\frac{4 \pi^2 e^2}{q^2}$. With this in mind, and to facilitate experimental comparisons, we define a scalar dielectric function by summing all the density-density components $\epsilon(q,i \nu_n) = \sum_{i,j} \epsilon^{GW + \mathrm{EDMFT}}_{ii,jj}(q,i \nu_n)$, analogously to Sect. \ref{chi}. We find that the inverse of this expression is causal and can be analytically continued with MaxEnt.

\subsection{Spin response function \label{spin_resp} }

One can probe the system's tendency towards magnetic order by studying its multi-orbital spin susceptibility $\chi_s$, which in the RPA formalism is defined as \cite{flex}
\begin{equation}
\chi_s(q,i\omega_n) = \chi_0(q,i\omega_n)[\mathbb{I} + U_s(i\omega_0) \chi_0(q,i\omega_n) ]^{-1}. 
\label{chi_s_eq}
\end{equation}
Here, the uncommon plus sign in the denominator originates from the minus sign in the definition of the charge susceptibility in Eq.~\eqref{charge_susc}.

The local spin interaction $U_s$ is obtained from the decomposition of the generalized multi-orbital Hubbard interaction on a given site,
\begin{eqnarray}
    H_{\mathrm{int}} = \frac{1}{4} \sum_{\alpha_1,...,\alpha_4} U^{(\alpha_1,\alpha_4),(\alpha_3,\alpha_2)} c^{\dag}_{\alpha_1}c^{\dag}_{\alpha_2} c^{}_{\alpha_3}c^{}_{\alpha_4},
\end{eqnarray}
into spin and charge channels:
\begin{eqnarray}
    U^{(\alpha_1 \alpha_4)(\alpha_3\alpha_2)} = -\frac{1}{2} U_s^{(l_1 l_4)(l_3 l_2)} \mathbf{\sigma}_{\sigma_1,\sigma_4} \cdot \mathbf{\sigma}_{\sigma_2,\sigma_3} + \frac{1}{2} U_c^{(l_1 l_4)(l_3 l_2)} \delta_{\sigma_1,\sigma_4} \delta_{\sigma_2,\sigma_3},
\end{eqnarray}
where $\alpha_i = (l_i,\sigma_i)$ combines orbital and spin degrees of freedom. For a Kanamori-type $t_{2g}$ Hamiltonian, the spin and charge interaction matrices read \cite{no_sc}
\begin{equation}
U^{(l l^{\prime})(n n^{\prime})}_{s,c}= \begin{cases}U & \left(l=l^{\prime}=n=n^{\prime}\right) \\
U', \, -U^{\prime}+2 J & \left(l=n \neq n^{\prime}=l^{\prime}\right) \\
J, \, 2 U^{\prime}-J & \left(l=l^{\prime} \neq n^{\prime}=n\right) \\
J' & \left(l=n^{\prime} \neq n=l^{\prime}\right)\end{cases},
\end{equation}
where $U$, $U'$, $J$, $J'$ are the conventional intra-orbital Coulomb, inter-orbital Coulomb, Hund's coupling and inter-orbital pair hopping interactions, respectively. 

Due to the multi-orbital nature of the model, we define a scalar spin susceptibility $\chi_s$ by summing over all the density-density components:
\begin{equation}
    \chi_s = \sum_{i,j} \chi_s^{(ii)(jj)}.
\end{equation}
We use the static value of the local bare impurity interaction $\mathcal{U}$ to approximate the values of $U$, $U'$ and $J$ to construct $U_s$, assuming $J'=J$. We note that directly inserting the obtained $U_s$ into Eq.~\eqref{chi_s_eq} results in unphysical values of $\chi_s$ due to an eigenvalue of $ -U_s \chi_0$ becoming greater than 1. To avoid this, it is necessary to rescale the spin interaction as $U_s \rightarrow \eta U_s$. A value of $\eta=0.1$ is used in all subsequent spin susceptibility plots. This renormalization is consistent with the strongly renormalized spin vertex found, for example, in the two-particle self-consistent approach \cite{vilk1997,simard2023}. Similar interaction tunings have been used before in RPA calculations  \cite{no_sc}. 

\begin{figure}
    \begin{centering}
\includegraphics[width=0.5\textwidth]{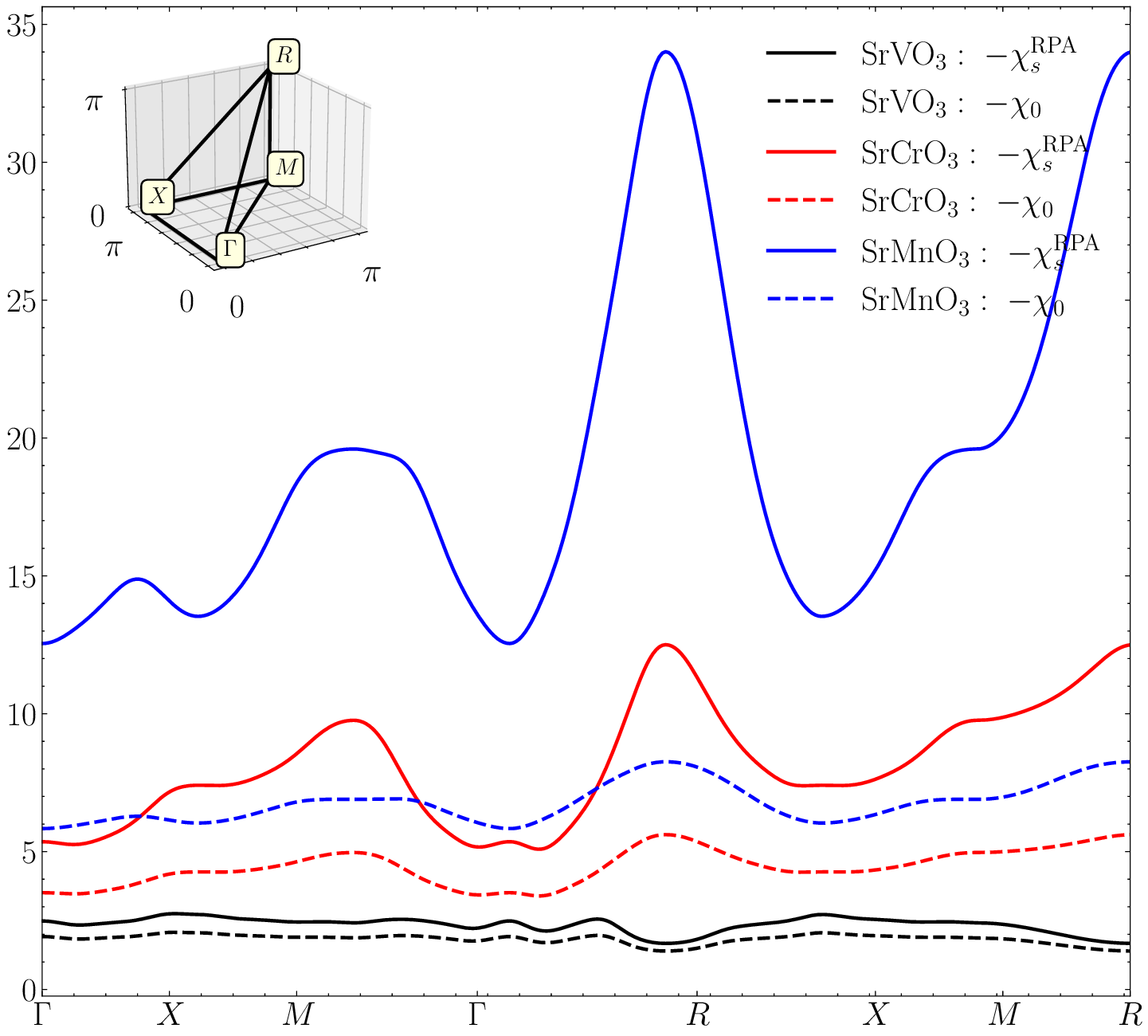}
\end{centering}
    \caption{Bare ($G^0G^0$) and RPA spin susceptibilities for the three cubic perovskites.}
    \label{chi_spin}
\end{figure}

\begin{figure}
    \begin{centering}
\includegraphics[width=\textwidth]{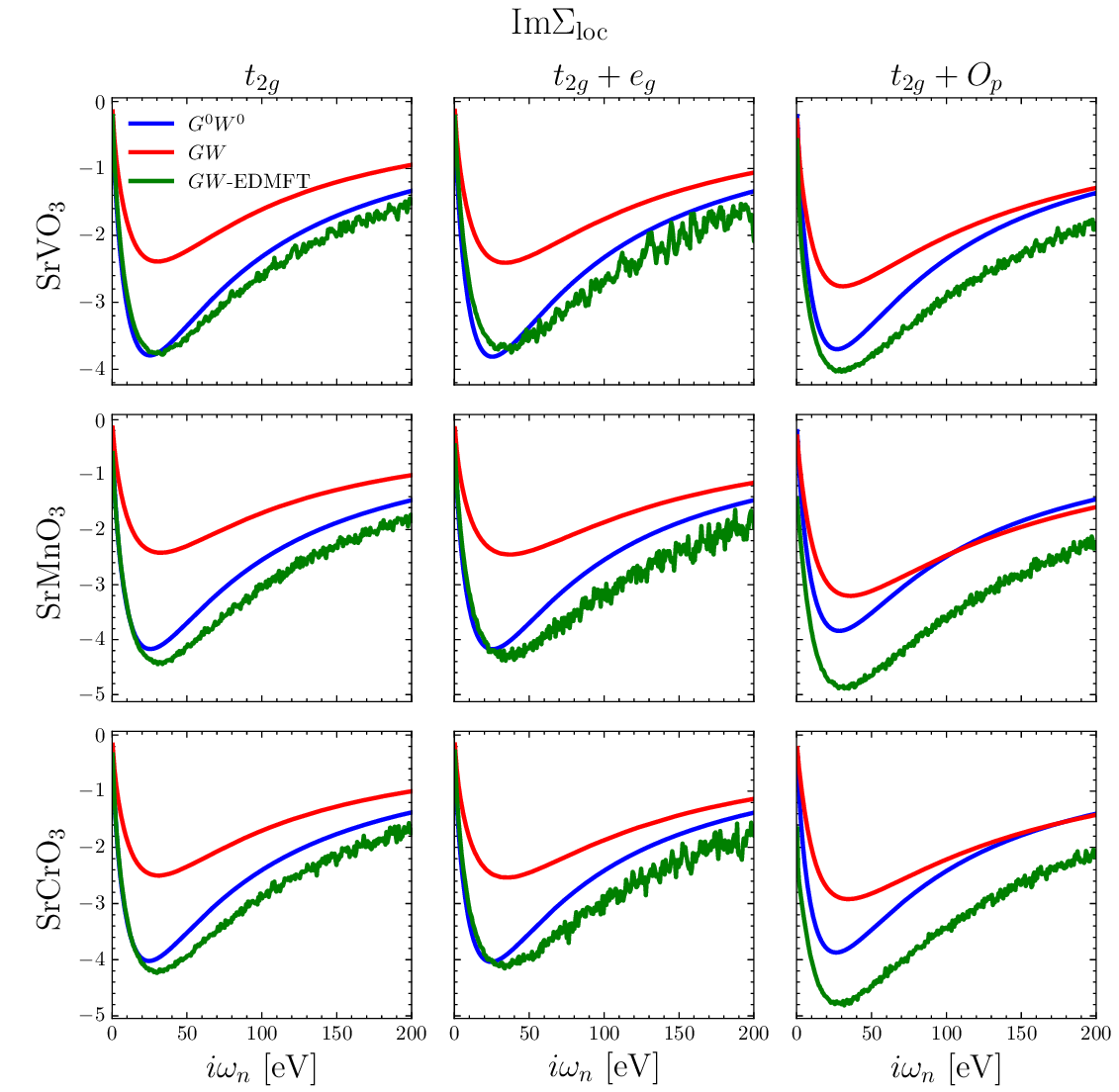}
\end{centering}
    \caption{Imaginary part of the local self-energy within different schemes for cubic perovskites in different model spaces.}
    \label{self_en_comparison}
\end{figure}

\newpage

\bibliography{srxo3_npj}